\documentclass[fleqn,usenatbib]{mnras}

\usepackage{newtxtext,newtxmath}

\usepackage[T1]{fontenc}
\usepackage{ae,aecompl}

\usepackage{hyperref}
\usepackage{graphicx}	
\usepackage{blindtext}  
\usepackage{natbib}
\usepackage{rotating,times,graphicx,latexsym}
\usepackage{color}
\usepackage{longtable}
\usepackage{lscape}
\usepackage{lipsum} 
\usepackage{array}
\usepackage{flafter}

\usepackage{orcidlink}

\usepackage{soul} 
\usepackage{xargs} 

\graphicspath{ {images_pdf/} }

\newcommand{\teff}{${\rm T}_{\rm eff}$}

\title[Spot properties on YSOs]{A survey for variable young stars with small telescopes - XI. Spot Lifetimes and Coverage Distributions}

\author[Dirk Froebrich et al.]{Dirk Froebrich\thanks{E-mail: df@kent.ac.uk}\orcidlink{0000-0003-4734-3345}\textsuperscript{1}, 
Carys Herbert\orcidlink{0000-0003-4217-8811}\textsuperscript{1,2},
Aleks Scholz\textsuperscript{3},
Benjamin W. Ryan\orcidlink{0009-0002-3759-3439}\textsuperscript{1},
Siegfried Vanaverbeke\orcidlink{0000-0003-0231-2676}\textsuperscript{4,5,6}, 
\newauthor 
Jochen Eisl\"offel\orcidlink{0000-0001-6496-0252}\textsuperscript{7}, 
Cledison Marcos da Silva\textsuperscript{8,9}, 
Michel Michaud\textsuperscript{10},
Kathy Griffiths\textsuperscript{11,8,12,13},
\newauthor 
Ari M. Siqueira\textsuperscript{11,8,9},
Thomas Urtly\textsuperscript{11},
Francisco C. Sold\'{a}n Alfaro\textsuperscript{14,8,15},
George Fleming\textsuperscript{11}, 
Gregg L. Ruppel\textsuperscript{16},
\newauthor 
Domenico Licchelli\textsuperscript{17,18},
Stephen R.L. Futcher\textsuperscript{11,8,19,20},
Ivan L. Walton\textsuperscript{11},
Matthias Kolb\textsuperscript{21,22},
\newauthor 
Mario Morales Aimar\textsuperscript{14,23}, 
Geoffrey Stone\orcidlink{0000-0001-5888-9162}\textsuperscript{8,24},
Yenal \"{O}{\u g}men\textsuperscript{25},
Simon Francis Dawes\textsuperscript{11,26},
\newauthor 
Franz-Josef Hambsch\orcidlink{0000-0003-0125-8700}\textsuperscript{5,8,21,27},
Nick J. Quinn\textsuperscript{11},
Juan-Luis González-Carballo\textsuperscript{28,14,8},
Franky Dubois\textsuperscript{4,5}, 
\newauthor 
Jos\'{e} Luis Salto Gonz\'{a}lez\textsuperscript{14,29,30},
Adam Popowicz\orcidlink{0000-0003-3184-5228}\textsuperscript{31},
Krzysztof Bernacki\orcidlink{0000-0003-4647-7114}\textsuperscript{31},
Slawomir Bednarz\orcidlink{0009-0000-1171-3370}\textsuperscript{31},
\newauthor 
Tim Nelson\textsuperscript{19,32},
Tony Vale\textsuperscript{11,8,12,33},
Tonny Vanmunster\textsuperscript{34,35},
Georg Piehler\textsuperscript{36},
Stephen C. Percy\textsuperscript{37},
\newauthor 
Jordi Berenguer Amatller\textsuperscript{14},
Jacopo Fallai\textsuperscript{38},
Amritanshu Vajpayee\orcidlink{0000-0003-0718-1708}\textsuperscript{39,40},
Faustino Garc\'{i}a de la Cuesta\textsuperscript{41},
\newauthor 
Rafael Castillo García\textsuperscript{14,42},
Dawid Mo\'{z}dzierski\textsuperscript{43},
Krzysztof Kotysz\textsuperscript{43,44},
Przemys{\l}aw J. Miko{\l}ajczyk\orcidlink{0000-0001-8916-8050}\textsuperscript{43,45},
\newauthor 
Holly Stokes-Geddes\textsuperscript{1},
Matthew D. Dickers\orcidlink{0000-0001-9615-9101}\textsuperscript{1},
Ivana I. Grozdanova\textsuperscript{1},
Chiara F. Kawohl\textsuperscript{1},
Aashini L. Patel\orcidlink{0000-0001-5800-6710}\textsuperscript{1},
\newauthor
James S. Urquhart\orcidlink{0000-0002-1605-8050}\textsuperscript{1},
Tim M. Kinnear\textsuperscript{1}
\\
\textsuperscript{1}Centre for Astrophysics and Planetary Science, School of Engineering, Mathematics and Physics, University of Kent, Canterbury CT2 7NH, UK\\
\textsuperscript{2}Royal Museums Greenwich, Romney Road, SE10 9NF, UK\\
\textsuperscript{3}SUPA, School of Physics \& Astronomy, University of St Andrews, North Haugh, St Andrews KY16 9SS, UK\\
\textsuperscript{4}Public observatory ASTROLAB IRIS, Provinciaal Domein “De Palingbeek”, Verbrandemolenstraat 5, 8902 Zillebeke, Ieper, Belgium\\
\textsuperscript{5}Vereniging Voor Sterrenkunde (VVS), Zeeweg 96, 8200 Brugge, Belgium\\
\textsuperscript{6}Centre for Mathematical Plasma-Astrophysics, Department of Mathematics, KU Leuven, Celestijnenlaan 200B, 3001 Heverlee, Belgium\\
\textsuperscript{7}Th\"uringer Landessternwarte, Sternwarte 5, D-07778 Tautenburg, Germany\\
\textsuperscript{8}American Association of Variable Star Observers (AAVSO), 185 Alewife Brook Parkway, Suite 410, Cambridge, MA 02138, USA\\ 
\textsuperscript{9}Variable Stars South (VSS), Royal Astronomical Society of New Zealand, PO Box 3181, Wellington, New Zealand\\
\textsuperscript{10}MCD Observatory, 23 Langlois, Saint-Anaclet, Quebec G0K1H0, Canada\\
\textsuperscript{11}British Astronomical Association, Variable Star Section, PO Box 702, Tonbridge, TN9 9TX, UK\\ 
\textsuperscript{12}Bath Astronomers, 19 New King Street, Bath BA1 2BL, UK\\
\textsuperscript{13}Flamsteed Astronomy Society, Royal Observatory Greenwich, Blackheath Avenue, London SE10 8XJ, UK\\
\textsuperscript{14}Observadores de Supernovas (ObSN), \href{https://www.obsn.es/}{\texttt{https://www.obsn.es/}}, Spain\\
\textsuperscript{15}Science Department, Seville University, Av. de la Ciudad Jard\'{i}n, 20-22, 41005 Sevilla, Spain\\
\textsuperscript{16}Dark Skies New Mexico Observatory, 1984 W. Golden Rose Pl, Oro Valley, AZ 85737, USA\\
\textsuperscript{17}R.P. Feynman Observatory, Piazzetta del Ges\'{u} 3, 73034, Gagliano del Capo, Italy\\
\textsuperscript{18}Center for Backyard Astrophysics (CBA), Piazzetta del Ges\'{u} 3, 73034, Gagliano del Capo,  Italy\\
\textsuperscript{19}Hampshire Astronomical Group, Hinton Manor Lane, Clanfield, PO8 0QR, UK\\
\textsuperscript{20}Royal Astronomical Society, Burlington House, Piccadilly, London W1J 0BQ, UK\\
\textsuperscript{21}Bundesdeutsche Arbeitsgemeinschaft für Ver\"{a}nderliche Sterne (BAV), Munsterdamm 90, 12169 Berlin, Germany\\
\textsuperscript{22}Sternwarte Neanderhoehe Hochdahl, Sedentaler Str. 105, 40699 Erkrath, Germany\\
\textsuperscript{23}Observatorio de Sencelles, Sonfred Road 1, 07140 Sencelles, Mallorca, Spain\\
\textsuperscript{24}Dimension Point Observatory, 14 Galaxy Point, Mayhill, New Mexico 88339, USA\\
\textsuperscript{25}Green Island Observatory, Karao{\u g}laono{\u g}lu Street 63A, Ge\c{c}itkale Ma{\u g}usa, North Cyprus\\ 
\textsuperscript{26}Crayford Manor House Astronomical Society, Parsonage Lane Pavilion, Parsonage Lane, Sutton-at-Hone, Dartford DA4 9HD, Kent, UK\\
\textsuperscript{27}Groupe Europ\'{e}en d’Observations Stellaires (GEOS), 23 Parc de Levesville, 28300 Bailleau l’Ev\^{e}que, France\\
\textsuperscript{28}Cerro del Viento Observatory (MPC~I84), Fernandez Pirfano Square 3, 06010 Badajoz, Spain\\
\textsuperscript{29}Sociedad Malague\~{n}a de Astronom\'{i}a (SMA), Centro Cultural Jos\'{e} Mar\'{i}a Guti\'{e}rrez Romero, Cl Rep\'{u}blica Argentina, no 9, Urb. El Limonar, 29016 M\'{a}laga, Spain\\
\textsuperscript{30}Cal Maciarol m\'{o}dul 8 Observatory, Masia Cal Maciarol, Cam\'{i} de l'Observatori S/N, 25691 \`{A}ger, Spain\\
\textsuperscript{31}Department of Electronics, Electrical Engineering and Microelectronics, Silesian University of Technology, Akademicka 16, 44-100 Gliwice, Poland\\
\textsuperscript{32}Horndean Observatory, 6 Falcon Road, Horndean, Waterlooville, Hampshire, PO89BY, UK\\
\textsuperscript{33}The Herschel Society, The Herschel Museum of Astronomy, 19 New King Street, Bath BA1 2BL, UK\\
\textsuperscript{34}Center for Backyard Astrophysics Extremadura, 06340 Fregenal de la Sierra, Spain\\
\textsuperscript{35}Vereniging voor Sterrenkunde VVS, 3401 Landen, Belgium\\
\textsuperscript{36}Selztal Observatory, D-55278 Friesenheim, Bechtolsheimer Weg 26, Germany\\
\textsuperscript{37}The Studios Observatory (MPC Z52), 31 Ipswich Gardens, Grantham NG31 8SE, UK\\
\textsuperscript{38}Osservatorio Conigliolo, Via Chiantigiana, Ginestra Fiorentina 50055, Firenze, Italy\\
\textsuperscript{39}Saptarishi India, 1/125-G, New Civil Lines, Navadiya, Fatehgarh, District - Farrukhabad, Uttar Pradesh, India - 209 601 \\
\textsuperscript{40}Akash Ganga: Centre for Astronomy, A/2, East \& West Villa, Nowroji Vakil Street, Grant Road, Mumbai, Maharashtra, India - 400 007\\
\textsuperscript{41}La Vara, Valdes Observatory( MPC~J38), Barrio La Bara, sin n\'{u}mero Mu\~{n}as de Arriba c\'{o}digo postal 33784, Asturias, Spain\\
\textsuperscript{42}Asociacion Astronomica Cruz del Norte, Calle Caceres 18, 28100 Alcobendas, Madrid, Spain\\
\textsuperscript{43}Astronomical Institute, University of Wroc{\l}aw, ul. M. Kopernika 11, 51-622 Wroc{\l}aw, Poland\\
\textsuperscript{44}Astronomical Observatory, University of Warsaw, Al. Ujazdowskie 4, 00-478 Warsaw, Poland\\
\textsuperscript{45}Astrophysics Division, National Centre for Nuclear Research, Pasteura 7, 02-093 Warsaw, Poland\\
}

\date{Accepted XXX. Received YYY; in original form ZZZ}

\pubyear{2024}

\begin{document}
\label{firstpage}
\pagerange{\pageref{firstpage}--\pageref{lastpage}}
\maketitle
\clearpage\newpage
\begin{abstract} 
We present a homogeneous analysis of rotational variability and spot properties in young stellar objects across multiple star-forming regions observed by the Hunting Outbursting Young Stars (HOYS) project. From over 2000 candidate members, we identify 144 YSOs with robust periodic signals and well-constrained multi-band amplitudes. The sample has a median age of $\sim$1~Myr, effective temperatures of 3500--6500~K (masses $\sim$0.6--2~M$_\odot$), and is dominated by Class~2 objects, one third of which exhibit inner disc dust emission. The rotation period distribution is strongly bimodal, with 55 percent fast rotators ($P<5.5$~d) and 45 percent slow rotators. Fast rotators are predominantly inner disc-less, whereas slow rotators include both disc-bearing and disc-free systems, indicating that disc braking alone cannot explain the observed rotational states. We derive spot properties from multi-band amplitudes and find that, after correcting for observational biases, the intrinsic cold-spot coverage distribution of fast rotators is well described by an exponential function. This implies that small spot coverages are intrinsically much more common than large ones, consistent with stochastic magnetic flux emergence governing spot formation. In contrast, slow rotators show a pronounced deficit of small cold spots. After considering observational biases and alternative physical explanations, we conclude that small spots on slowly rotating YSOs have significantly shorter lifetimes. These results provide new evidence that magnetic surface structure and its evolution depend on stellar rotation, placing new empirical constraints on models of magnetic activity and angular momentum evolution in young stars.
\end{abstract}




\section{Introduction}

Stellar rotation is a key parameter in early stellar evolution, governing angular momentum redistribution, magnetic activity, and star–disc interactions. Observations of young stellar objects (YSOs) show a wide dispersion in rotation periods and associated variability, indicating complex regulation mechanisms. Variability on timescales of days is typically linked to structures within the co-rotation radius or on the stellar surface \citep{2014AJ....147...82C, 2021MNRAS.506.5989F, 2023MNRAS.520.5433H}, while longer-term variability can arise from occultations by circumstellar material on timescales from days to years \citep[e.g.][]{1989MNRAS.239..665L, 2023ASPC..534..355F}, or changes in long-term accretion rates \citep[e.g.][]{2014prpl.conf..387A}. The rotation of a young star is initially inherited from its parent molecular cloud, yet by the main sequence its angular momentum is reduced by $\sim$6–7 orders of magnitude \citep{2013EAS....62...25B}. During the first few Myr, rotation rates remain approximately constant despite ongoing accretion, implying efficient angular momentum loss, after which stars spin up as they contract towards the zero-age main sequence, reaching periods as short as $\sim$0.2–0.3~d, before spinning down again due to magnetised winds \citep{2007prpl.conf.....R, 2014prpl.conf..433B}. Thus, the observed period distribution of YSOs is bimodal \citep[e.g.][]{2002A&A...396..513H, 2007prpl.conf..297H, 2009A&A...502..883R, 2010A&A...515A..13R, 2023RAA....23g5015W}, indicating two distinct rotational states and suggesting that the transition between them occurs on relatively short timescales.

A widely adopted explanation for this behaviour is disc braking, in which magnetic coupling between the star and its protoplanetary disc regulates stellar rotation by extracting angular momentum. This can be achieved in various ways, e.g. by disc locking \citep{1991ApJ...370L..39K}, or accretion powered winds \citep{2005ApJ...632L.135M}. Observational evidence supports a link between slow rotation and accretion, particularly for stars in the mass range $0.4$–$1.2$~M$_\odot$ \citep{2006ApJ...647L.155F} and suggests that the magnetically coupled region is confined to the vicinity of the co-rotation radius \citep{2013A&A...550A..99Z}. However, the efficiency of this mechanism and its ability to reproduce observed rotation distributions remain uncertain. The development of a radiative core on timescales of $0.5$–$10$~Myr can alter the stellar magnetic field topology, weakening the large-scale dipolar component required for effective disc braking \citep{2019A&A...632A...6G}. Moreover, some studies find no statistically significant correlation between accretion indicators and rotation period \citep{2006AJ....132.2135S}, and the coexistence of fast rotators with discs and slow rotators without discs indicates that additional processes, such as magnetic field evolution, binarity, or rapid disc dispersal, must play an important role. Together, these results suggest that while disc braking provides a useful framework, it does not fully account for the diversity of rotational properties observed in YSOs.

Cold surface spots on YSOs are the manifestation of strong stellar magnetic fields that inhibit convection, analogous to sunspots but on a much larger scale. Solar observations provide a useful reference framework: sunspots trace a cyclic magnetic field with well-defined polarity patterns \citep{1919ApJ....49..153H}, although a small fraction deviate from this behaviour \citep{2018ApJ...867...89L}. In YSOs, magnetic field strengths reach a few kilogauss \citep{2007ApJ...664..975J}, significantly exceeding solar values \citep{1961ApJ...133..572B}, leading to larger, more stable spots that can cover tens of percent of the stellar surface. These spots exhibit temperature contrasts of several hundred to $\sim$1400~K below the photosphere \citep{2008A&A...479..827G} and their overall coverage decreases with stellar age \citep{2020AJ....160....5M}. Photometric monitoring demonstrates that spot modulation is a dominant source of variability, and most YSOs show signatures consistent with cold spot coverage \citep{1999A&A...349..619B, 2001AJ....121.3160C}.

Hot spots arise from magnetospheric accretion, where disc material is channelled along magnetic field lines onto the stellar surface, producing localized heating and variability. These spots drive a range of observed light-curve morphologies, from burst-like behaviour linked to episodic accretion \citep{2017ApJ...836...41C, 2016AJ....151...60S} to stochastic variability caused by unstable accretion flows \citep{2016AJ....151...60S}, and  quasi-periodic modulation reflecting rotation combined with evolving accretion structures \citep{2014AJ....147...82C}. Contrary to the classical picture of small, very hot accretion shocks, observations reveal a wide diversity in spot properties: temperature contrasts can be only a few hundred Kelvin above the photosphere with large filling factors up to $\sim$40~\% \citep[e.g.][]{2012MNRAS.419.1271S, 2016MNRAS.458.3118B, 2023MNRAS.526.4885S}. In other cases hot spots occupy less than 1~\% of the surface \citep{2013ApJ...767..112I}. Extended, low-density accretion structures can also dominate optical emission \citep{2021Natur.597...41E}, highlighting the complexity of accretion-driven variability in YSOs.

The Hunting Outbursting Young Stars (HOYS) citizen science project combines long-term monitoring (from multi-year to decade-long) with multi-band optical photometry in standard filters, from a coordinated network of amateur and professional observers that enables high-cadence homogeneous coverage across multiple regions \citep{2018MNRAS.478.5091F, 2020MNRAS.493..184E}. Thus, it provides a uniquely suited dataset for investigating the evolution of rotational variability in YSOs. Previous HOYS studies have demonstrated its capability to characterise rotational variability and surface features in detail. For example, individual objects such as V1598~Cyg exhibit rapid rotation and spot driven variability \citep{2020MNRAS.497.4602F}, while systematic analyses in IC~5070 revealed a clearly bimodal period distribution and allowed the derivation of spot temperatures and surface coverage for sizeable samples of YSOs \citep{2021MNRAS.506.5989F, 2023MNRAS.520.5433H, 2024MNRAS.529.4856H}. These works show that most variability is dominated by cold spots with large filling factors, along with a smaller population of warm or hot spots linked to accretion, but are limited to individual regions and relatively modest sample sizes.

The primary motivation of this study is to extend this analysis across multiple star-forming regions within the HOYS dataset, thereby testing whether key observational features, such as bimodal period distributions, spot properties, and the connection between rotation and circumstellar discs, are universal or environment-dependent. Expanding the sample enables a statistically robust characterisation of rotational and spot properties, while consistent period analysis across epochs provides constraints on spot lifetimes and on the intrinsic distribution of spot properties. By linking rotation periods, variability amplitudes, and disc indicators, this work aims to assess the role of magnetic star–disc interaction in regulating angular momentum and to disentangle the relative contributions of magnetic activity, accretion, and circumstellar material in shaping YSO variability.

This paper is structured as follows: In Sect.~\ref{data} we detail the extraction and calibration of our HOYS data, the period and effective temperature determination, as well as the spot property calculations. The period distribution and the YSO properties are discussed in Sect.~\ref{yso_sample}. In Sect.~\ref{discussion} we discuss how our results compare to previous work, and what the implications of the spot property distributions are for the nature of the cold and warm spots on YSOs.

\section{Data and Methodology}\label{data}

In this section, we provide a brief overview of the data and methodology used. They are mostly identical to what has been done in our previous work in IC~5070 \citep{2021MNRAS.506.5989F, 2023MNRAS.520.5433H, 2024MNRAS.529.4856H}, but they are applied here to all HOYS fields. Any differences are detailed in the following.

\subsection{Data, Calibration, and Object Selection}

All photometry data used are from the HOYS citizen science project \citep{2018MNRAS.478.5091F}, which monitors 25 young (age less than 10~Myr) and nearby (distance less than 1~kpc) star clusters and star forming regions with a combination of amateur, university, and professional telescopes. The aim of the project is to obtain daily images in optical broad band filters over about 25~yr. The images are calibrated relative to a set of reference images taken under photometric conditions in the $V$, $R_C$, and $I_C$ filters. These will be referred to as $V$, $R$, and $I$, throughout, for simplicity. Data in other filters ($U$, $B$, $H_\alpha$) are available but not used in this work. All data are colour corrected (in case of non-ideal observing conditions or non-standard filters) following the procedures established in \citet{2020MNRAS.493..184E}.

In \citet{2024MNRAS.529.1283F} we used Gaia~DR3 data \citep{2016A&A...595A...1G,2023A&A...674A...1G} to generate a full list of cluster members in all HOYS target regions based on their astrometric properties (parallax and proper motion). These objects therefore have distances and space motions consistent with membership of the respective young star-forming regions, making contamination by unrelated foreground or background stars, including evolved giants, very small. In that work a total of just over 3000 objects were selected as potential HOYS cluster members for investigation. Their light curves were extracted from our data base on 21 October 2022 and calibrated using the procedures detailed above. Light curves with at least 100 data points in each of the three filters are available for 2047 of these. Only those are investigated in this work.

For all potential cluster members, we used Gaia~DR3 \citep{2016A&A...595A...1G,2023A&A...674A...1G} cross-matches to the Two Micron All-Sky Survey \citep[2MASS,][]{2006AJ....131.1163S} and the Wide-Field Infrared Survey Explorer All-sky catalogue \citep[WISE,][]{2013yCat.2328....0C} to obtain their near- ($J$, $H$, $K$) and mid-infrared ($W1$, $W2$, $W3$, $W4$) magnitudes and colours. In some cases, the objects did not have any $K$-band data available in 2MASS. For these we extracted data from the UK Infrared Deep Sky Survey \citep[UKIDSS,][]{2007MNRAS.379.1599L}, in particular from the Galactic Plane Survey \citep[GPS,][]{2008MNRAS.391..136L}.

A note on notation: Throughout the paper we are investigating the light curves, or parts of them, for our objects in multiple filters ($V$, $R$, and $I$). In some cases, we will only need to refer to data in an individual filter, and in others to the entire set of data in all filters. Thus, for simplicity and to maintain the notation used in \citet{2024MNRAS.529.4856H}, we will denote these data sets using $\{ \}$ around the symbol for the shortest wavelength filter included in the set. For example, $\{ V \}$ refers to the data set in the filters $V$, $R$, and $I$. 

\subsection{Light Curve Slicing}

We plan to investigate the spot properties on the observed YSOs over time. For this purpose, all light curves have been dissected (sliced) into blocks of six months length. This has been done with an oversampling of two. In other words, a six month long slice is created every three months for all light curves. This follows the procedure developed for the field IC~5070 in \citet{2024MNRAS.529.4856H}. The earliest data point in any of our light curves is from 2013. For each object, the starting date for the first slice is therefore set on the date of its worst observability in the Northern Hemisphere in 2013. This is the date when it has the lowest altitude at midnight. 

For all objects, only slices that contained at least 50 data points in each of $\{V\}$ were retained. The majority of light curves did not have the necessary observational cadence to meet this threshold for their entirety. From the 2047 light curves, 918 objects had at least one slice with sufficient data. In total, there are 3752 individual, 6~month long slices. Table~\ref{regions} lists the HOYS target regions (including their number and nominal coordinates) with investigated light curves, their number, and the number of slices with sufficient data. Unlisted HOYS regions do not have any light curves with sufficient data.

\begin{table}
\caption{\label{regions} HOYS target regions investigated for periodic YSOs. We list the region name, the three digit region ID, the nominal centre coordinates for the fields, the number of light curves N$_{\rm LC}$ and slices N$_{\rm SL}$ investigated with sufficient data, and the number of light curves and slices found to show a periodic signal.}
\centering
\setlength{\tabcolsep}{3.7pt}
\renewcommand{\arraystretch}{1.0}
\begin{tabular} {|c|c|c|c|c|c|c|c|}
\hline
\multicolumn{2}{|c|}{Region} & \multicolumn{2}{c|}{RA/DEC (J2000)} & \multicolumn{2}{c|}{Sufficient Data} & \multicolumn{2}{c|}{Periodic} \\ \hline
Name & ID & h:m:s & $^\circ$:':" & N$_{\rm LC}$ & N$_{\rm SL}$ & N$_{\rm LC}$ & N$_{\rm SL}$ \\ \hline
NGC~1333      & 001 & 03:29:02 & $+$31:20:54 & 20 & - & - & - \\
IC~348        & 002 & 03:44:34 & $+$32:09:48 & 79 & - & - & - \\
M~42          & 003 & 05:35:17 & $-$05:23:28 & 74 & 155 & 9 & 34 \\
NGC~2264      & 004 & 06:40:58 & $+$09:53:42 & 285 & 1 & - & - \\
$\sigma$~Ori  & 005 & 05:38:45 & $-$02:36:00 & 69 & 324 & 17 & 90 \\
$\lambda$~Ori & 006 & 05:35:06 & $+$09:56:00 & 18 & - & - & - \\
L~1641~N      & 007 & 05:36:19 & $-$06:22:12 & 108 & 130 & 11 & 22 \\
NGC~2244      & 008 & 06:31:55 & $+$04:56:30 & 58 & - & - & - \\
Berkeley~86   & 110 & 20:20:12 & $+$38:41:24 & 229 & 25 & - & - \\
P~Cyg         & 111 & 20:17:47 & $+$38:01:59 & 196 & 15 & - & - \\
IC~1396~N     & 114 & 21:40:42 & $+$58:16:06 & 150 & 405 & 17 & 71 \\
IC~1396~A     & 115 & 21:36:35 & $+$57:30:36 & 187 & 528 & 25 & 78 \\
NGC~7129      & 116 & 21:42:56 & $+$66:06:12 & 25 & 24 & 1 & 4 \\
IC~5070       & 118 & 20:51:00 & $+$44:22:00 & 136 & 970 & 31 & 211 \\
IC~5146       & 119 & 21:53:29 & $+$47:16:01 & 65 & 121 & 9 & 19 \\
V898~Ori      & 202 & 05:40:26 & $-$07:05:37 & 13 & 30 & 3 & 8 \\
YY~Ori        & 203 & 05:34:48 & $-$05:57:57 & 52 & 160 & 9 & 32 \\
V555~Ori      & 210 & 05:35:09 & $-$04:46:52 & 80 & 226 & 10 & 48 \\
Gaia~19~fct   & 213 & 07:09:21 & $-$10:29:35 & 145 & 409 & 2 & 3 \\
Gaia~19~eyy   & 214 & 08:30:42 & $-$41:33:43 & 58 & 229 & - & - \\
\hline
\end{tabular}
\end{table}

\subsection{Period Search and Amplitude determination}\label{per_amp}

The search for periodic variability has been conducted following our investigations in \citet{2021MNRAS.506.5989F}. In this work, we established that the most efficient way of searching for periodic signals in in-homogeneously sampled light curves such as those from HOYS is a combination of different methods. In particular, we combine the results of the L1Boot, L1Beta, L2Beta, and GLS periodogram methods. The details of these are discussed in the Appendix of \citet{2021MNRAS.506.5989F}. We follow the procedures for period determination in each slice and object, as outlined in \citet{2024MNRAS.529.4856H} for the IC~5070 region, with some minor adjustments.

For each object, only slices with 50 data points in all $\{V\}$ filters were considered, since the amplitudes in all three filters are required to determine the spot properties later on. Periodograms with 1000 test periods, homogeneously sampled in frequency between 1/0.5~d$^{-1}$ and 1/20~d$^{-1}$ are determined for each filter, periodogram method, and slice. Candidate periods for a slice were accepted if the maximum periodogram power exceeded 0.15 (for L1Boot, L1Beta, L2Beta) or the lowest False Alarm Probability (FAP) was below 0.1~\% (for GLS). Each of these candidate periods must be within 5~\% of each other in at least two of the three filters by the same method. All candidate periods within one percent of 0.5~d, 1~d, or 20~d were removed, as they are either at the edges of the investigated parameter space or too close to the typical observing cadence. This generates a list of candidate periods for each object.

These lists can contain different periods. In principle, for a given object, we can have four candidate periods per slice, one from each of the methods. Thus, we clustered the list of all candidate periods per object to identify groups. If the most populous group contained more than 2/3$^{\rm rd}$ of all candidate periods (and there are at least five in the group), the median of all periods in that group was chosen as the candidate period for the object. If for an object none of the groups had sufficient data, the phase folded light curves were inspected manually, and visually good quality periods were added. The minimum number of candidate periods accepted was two, which were detected by two different methods in a single slice.

Finally, for each object with a candidate period, the final period was determined as follows. A simple Lomb-Scargle periodogram is applied over the entire light curve in each filter, centred on the median of the candidate periods in the primary group. We used 1000 homogeneously distributed test periods within 10 percent of that median. The period associated with the highest peak in the periodogram was determined in each filter. The median of these is taken as the final period for the object, and the standard deviation as the uncertainty. The full list of all identified periodic sources is shown in Table~\ref{rotationperiods}  in the Appendix. In that table, we list the Gaia~DR3 ID, the Right Ascension and Declination, as well as the measured period and the uncertainty.

All slices of the periodic variables with sufficient data were phase folded with the respective final period for each object. The observed sets of peak-to-peak amplitudes $\left( \hat{A}^o_{\{V\}} \right)$ were calculated following \citet{2024MNRAS.529.4856H}. These were taken from the running median in the light curve phase plot, smoothed over 0.1 in phase. The associated uncertainties are the standard deviation of the phase folded light curve from the smoothed running median. In addition, the phase position of the minima and maxima of the smoothed running median was recorded. Only slices where the signal-to-noise ratio for the peak-to-peak amplitudes in all three filters was greater than three and where minima and maxima in the phase plots aligned within 0.125 are considered for spot fitting. All phase folded light curves were inspected, and five additional slices were removed because the running median did not represent the data well. This is generally due to gaps in the data in phase space, which commonly occur with periods close to integer days. In total, 620 sets of amplitudes $\hat{A}^o_{\{V\}}$ in 144 YSOs were identified as reliable. Table~\ref{regions} lists their numbers for each of the HOYS regions. 

\subsection{Effective Temperature Estimates and Spot Fitting}\label{spotfitting}

The fitting of spot properties on the YSOs follows the procedure outlined in \citet{2023MNRAS.520.5433H}. This only requires knowledge of the effective temperature \teff\ of the star and the sets of amplitudes from the spot induced periodic variability. In our previous analyses of spot properties \citep{2023MNRAS.520.5433H,2024MNRAS.529.4856H} we focused on the YSOs in the IC~5070 region. There, almost all objects had effective temperatures available from \citet{2020ApJ...904..146F}. For sources without, we used the best estimates based on their position in the Gaia colour-magnitude diagram. Unfortunately, this is not an option for most other HOYS regions. There are no consistent catalogues of effective temperatures for all YSOs in these fields available. As demonstrated in \citet{2023MNRAS.520.5433H}, systematic changes of up to $\pm400$ K in the adopted \teff\ values do not alter the inferred spot property distributions beyond the statistical uncertainties. Thus, we outline below our new method to obtain approximate effective temperatures from survey broad band colours for all HOYS YSOs.

We extracted all Gaia members in the North American and Pelican Nebula (NAP) region that have an effective temperature listed in \citet{2020ApJ...904..146F}. This list was limited to \teff\ $< 8000$~K. In addition, we extracted all Gaia~DR3 magnitudes ($G, BP, RP$) and parallax values. The $G$-band magnitude was converted into an absolute magnitude $G_G$ using the parallax, without considering any extinction. In addition, all 2MASS and WISE magnitudes (if available) were obtained. Homogeneous and reliable extinction estimates are not available for the full calibration sample. This leaves a list of $\sim$200 members of the NAP cluster with effective temperatures in \citet{2020ApJ...904..146F}. 

In order to fit the effective temperature in this sample based on their absolute $G$-band magnitude and different colours, we define the function shown in Equ.\ref{tefffit}. There, the symbols take their usual meaning and the $p_i$ are free parameters. The two additional colours $C_1$ and $C_2$ represent any colour that can be created from the magnitudes of Gaia ($G, BP, RP$), 2MASS/UKIDSS GPS ($J, H, K$), and WISE ($W1, W2$). We do not consider $W3$ or $W4$ because these are not available at a high signal to noise ratio for most objects. 

\begin{equation}
\begin{split}
{\rm T}_{\rm eff} = &  p_0 + p_1 \cdot G_G + p_2 \cdot (BP - RP) + ... \\
& \quad ... + p_3 \cdot G_G \cdot (BP - RP) + p_4 \cdot C_1 + p_5 \cdot C_2
\label{tefffit}
\end{split}
\end{equation}

\begin{figure*}
\centering
\includegraphics[angle=0,width=1.015\columnwidth,alt={Histogram comparing effective temperature distributions for three samples of young stellar objects. The blue outlined sample has a strong concentration near 3900 to 4000~K, with smaller numbers extending from about 3000 to 6600~K. The red filled sample is broader, peaking between roughly 4200 and 4800~K and extending to about 6700~K. The green filled sample contains fewer sources overall and is mostly concentrated between about 3800 and 5000~K, with a few hotter sources. The figure shows that the North America and Pelican Nebula young stellar objects are generally concentrated at cooler temperatures, while the larger rotational-variable sample spans a wider temperature range.}]{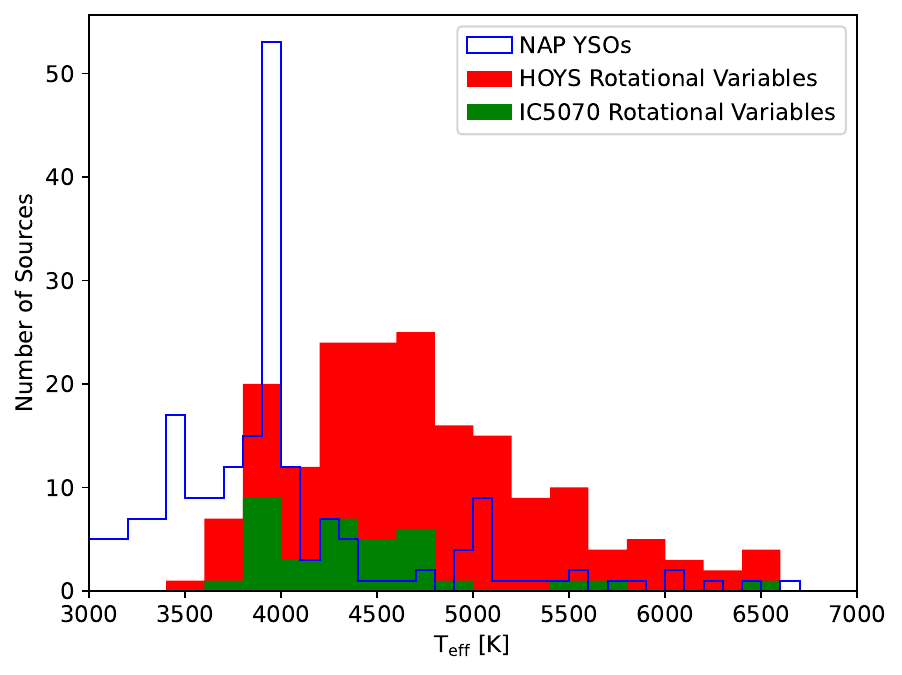} \hfill
\includegraphics[angle=0,width=0.985\columnwidth,alt={Colour–magnitude diagram of the HOYS sample showing Gaia $BP−RP$ colour against $G$-band magnitude. Points are coloured by effective temperature, with hotter sources shown in blue to purple and cooler sources shown in yellow to red. The hotter objects are mostly located at $BP−RP$ values between about 1 and 2 and $G$ magnitudes between about 3 and 5, while the cooler objects are generally redder, with $BP−RP$ greater than about 2 and $G$ magnitudes between about 6 and 8. One million year and four million year isochrones are overplotted as solid and dashed black lines, respectively. Many of the sources lie between these two isochrones, with some objects lying below the one million year track and others above the four million year track. A reddening vector corresponding to AV=1 mag is shown in the upper-right region of the plot, indicating the direction in which extinction would move sources in the diagram. Overall the distribution is consistent with a predominantly young stellar population with ages less than four million years.}]{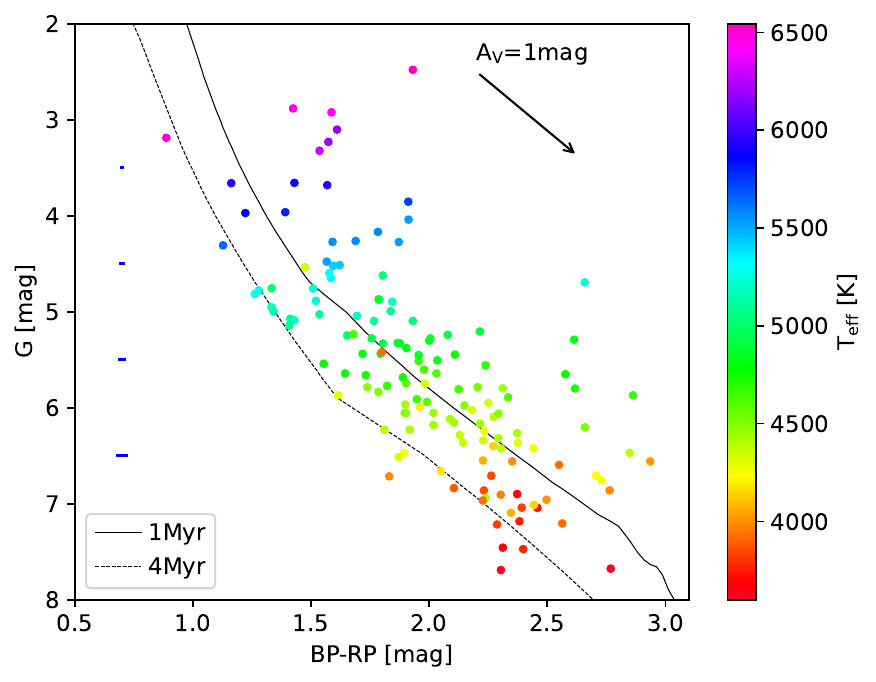} 
\caption{\label{teff_fit} {\bf Left:} In blue we show the \teff\ distribution of the YSOs in the NAP region from \citet{2020ApJ...904..146F}. In red the distribution of the fitted \teff\ values for the rotational variables in all HOYS fields is displayed, while in green the subsample of sources in IC~5070 is shown. {\bf Right:} A Gaia colour vs. absolute Gmag diagram of the rotational variables in our sample. The symbols are colour coded with the fitted \teff\ values. Typical Gaia colour errors are indicated as blue lines on the left hand side. An extinction vector and 1~Myr and 4~Myr PARSEC isochrones from \citet{2012MNRAS.427..127B} are over plotted for reference.}
\end{figure*}

The \teff\ distribution (blue histogram in the left panel of Fig.\ref{teff_fit}) of the \citet{2020ApJ...904..146F} NAP sample shows unusual peaks near 3500, 4000, and 5000~K, where objects appear to pile up, suggesting potential issues with these sources. We therefore fitted \teff\ with and without these objects. Furthermore, because the sample contains far fewer high-temperature than low-temperature sources, we also performed the fitting with and without weighting each source (e.g. by 1/\teff).

We used SciPy \citep{2020zndo...4406806V} to obtain the best-fitting (lowest RMS) parameters $p_i$ for every combination of the two colours $C_1$ and $C_2$, with and without weighting and removal of objects in the \teff\ peaks. All combinations yield very similar RMS and \teff\ values, with the worst RMS only 10 percent higher than the best. We therefore used all sources from \citet{2020ApJ...904..146F} without weighting, adopting $C_1 = K - W2$ and $C_2 = H - W1$ to fit \teff. These colours produce the lowest RMS among those available for all investigated periodic variables.

The adopted colour combination includes the WISE bands, raising the possibility that infrared excess from circumstellar discs could bias the fitted \teff\ values. However, testing all available colour combinations yields consistent temperatures within the uncertainties, with no systematic differences between objects with and without infrared excess (see Sect.~\ref{innerdisc}). Likewise, although simultaneous fitting of extinction and \teff\ would in principle be desirable, homogeneous and reliable extinction estimates are not available for the full sample. Since our calibration is empirical and based on spectroscopically determined \teff\ values, any average reddening present in the calibration sample is implicitly incorporated into the fit. Moreover, as demonstrated in \citet{2023MNRAS.520.5433H}, systematic changes of several hundred Kelvin in \teff\ alter the inferred spot temperatures by a comparable amount but have negligible effect on the spot coverages and do not change the distributions or conclusions presented here.

We show the distribution of the fitted values of \teff\ for our sample of rotational variables as a red histogram in the left panel of Fig.~\ref{teff_fit} and list the values for all periodic sources in Table~\ref{rotationperiods} in the Appendix. The \teff\ distribution is much smoother than that of the \citet{2020ApJ...904..146F} sample and has no significant narrow peaks. It has a maximum at temperatures of about 4500~K, a much smaller fraction of objects below 4000~K, and a large population of higher temperature sources. This distribution is understandable in the context of the way these objects have been selected. They are from a mix of all HOYS target regions, which are at various distances of up to 1~kpc. The sources need to be sufficiently bright in the $V$, $R$, and $I$-band filters for their period and amplitudes to be measured. Thus, this sample is dominated by brighter, more massive stars than the NAP sample from \citet{2020ApJ...904..146F}. For comparison, we over-plotted the \teff\ distribution for the sample of periodic variables in IC~5070. This accounts for about one fifth of the entire HOYS sample. 

The right hand panel of Fig.~\ref{teff_fit} shows the entire sample in a Gaia colour vs absolute magnitude diagram, where the symbols are colour coded based on the \teff\ values. We also plot 1~My and 4~Myr PARSEC isochrones \citep{2012MNRAS.427..127B} for reference, as well as a reddening vector. We can see that the fitted values of \teff\ generally follow the expected trend with some outliers. Comparison with the over plotted PARSEC isochrones indicates that the sample of periodic variables consists predominantly of stars younger than about 4~Myr, with roughly half the sources lying above and half below the~1 Myr isochrone, implying a median age of approximately 1~Myr.  Several sources are significantly above the 1~Myr isochrone. Thus, these are likely to be very young sources, unresolved higher-order multiples, objects affected by local nebulosity, or highly reddened objects.

The fitted values of \teff\ and the observed sets of $\hat{A}^o_{\{V\}}$ values are used for each object and slice to fit the spot temperature and coverage with their uncertainties. These follow exactly the procedures outlined in \citet{2023MNRAS.520.5433H,2024MNRAS.529.4856H}. The only change made is that we extended the maximum spot temperature considered from 12000~K to 14000~K. This was done due to some slightly higher effective temperature objects in our sample compared to our previous work in the IC~5070 field \citep{2024MNRAS.529.4856H}. As demonstrated in Appendix~A of \citet{2023MNRAS.520.5433H}, systematic changes of a few hundred Kelvin in the adopted stellar \teff\ result in approximately half that change in the fitted spot temperature, while the inferred spot coverage changes by only a few per cent. These variations are well within the typical uncertainties of the spot fitting (about 300~K for the spot temperature and 15 per cent for the spot coverage) and therefore do not affect the spot property distributions discussed in this paper.

\begin{figure}
\centering
\includegraphics[angle=0,width=\columnwidth,alt={A histogram showing the period distribution of two samples of rotational variables. The entire HOYS sample is shown in red and the IC~5070 field shown in green. Both samples show a bi-modal distribution with periods just under four and eight days with the shorter period mode having a higher peak. A CDF is plotted for both samples, solid for the HOYS sample and dashed for the IC~5070 sample. These are similar for both samples in shape and consistent with bi-modal distributions. There is a vertical dotted line at approximately 5.5~d separating slow and fast rotators. The figure shows that rotational variables are split between slow and fast rotators that speed up quickly due to the dip in number of objects between the two modes.}]{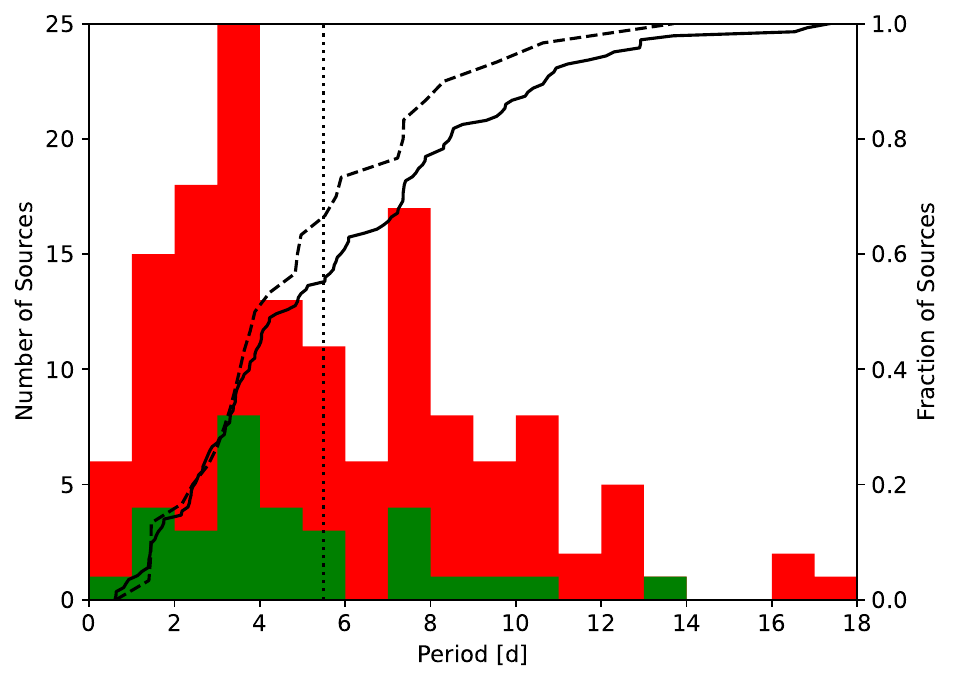} 
\caption{\label{Per_hist} The period distribution of all rotational variables identified in the HOYS target fields is shown as a red histogram. The corresponding CDF is shown as a solid black line. The green histogram and the dashed CDF are the data for the objects in the IC~5070 field. The vertical dotted line separates fast and slow rotators.} 
\end{figure}

\section{Rotational YSO Sample Characterisation}\label{yso_sample}

In this section, we discuss the stellar properties of the sample of rotational variables identified in all HOYS fields. 

\subsection{Period Distribution}\label{per_dis}

As shown in Sect.~\ref{per_amp}, we have reliably measured periods and amplitudes for 144 objects. The period distribution and the cumulative distribution function (CDF) are shown in Fig.~\ref{Per_hist}. The distribution is clearly bimodal, in agreement with previous studies of solar-type stars \citep[e.g.][]{2002A&A...396..513H, 2007prpl.conf..297H, 2009A&A...502..883R, 2010A&A...515A..13R, 2023RAA....23g5015W}. The CDF suggests that the gap between the two peaks lies between six and seven days. For consistency with our previous work \citep{2021MNRAS.506.5989F, 2023MNRAS.520.5433H, 2024MNRAS.529.4856H}, we adopt 5.5~d as the boundary between fast and slow rotators. This yields 79 (55 percent) fast rotators and 65 (45 percent) slow rotators. Modest changes to this boundary do not significantly affect these fractions.

\begin{figure}
\centering
\includegraphics[angle=0,width=\columnwidth,alt={A heat map of simulated average detection rate in our data as described in Sect.~3.1 as a function of period and $V$-band magnitude with the colorbar denoting detection rate. Up to 0.05~mag the detection rate remains at around 0.3, evenly for all periods above one day. The detection rate increases as amplitude increases evenly above a one day period up to a maximum detection rate of 0.75 at an amplitude of 0.16~mag. There is a consistently higher detection rate for periods less than one day. The figure shows that there is no bias in the period distribution above one day due to detection rate. }]{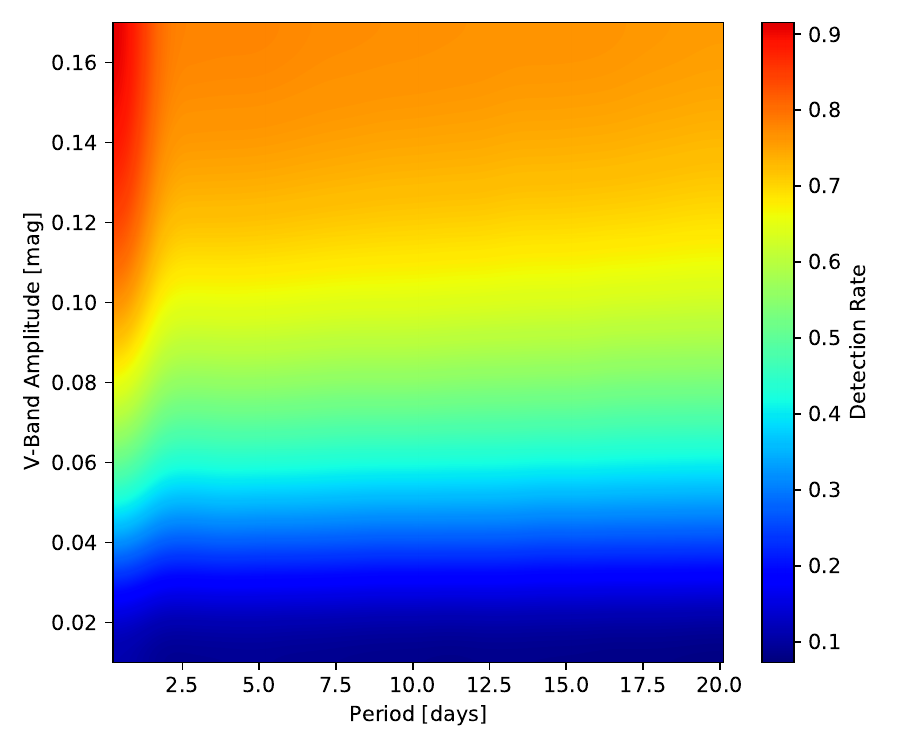} 
\caption{\label{Per_det_rate} Simulated average detection rate in our data as described in Sect.~\ref{per_dis} as a function of period and $V$-band amplitude. }
\end{figure}

To assess potential biases in period detection, we performed injection–recovery simulations using our data. Artificial sinusoidal signals with periods between 0.75 and 19.6~d (in steps of 1.05~d) and $V$-band amplitudes from 0.02 to 0.16~mag (in steps of 0.02~mag) were added to all slices with sufficient data. The corresponding amplitudes in $R$ and $I$ were scaled by factors of 0.87 and 0.71, respectively, following \citet{2023MNRAS.520.5433H}. The signals were injected repeatedly with random phase and the detection rate was recorded. As shown in Fig.~\ref{Per_det_rate}, detection rates averaged over all objects depend primarily on amplitude and only weakly on period, decreasing by at most a few percent across the period range considered. This reflects the fact that periods up to 20~d are sampled within $\sim$180~d windows, so the reduced number of cycles at longer periods has only a minor impact on detectability.

Since the period distribution combines multiple HOYS regions, we examine variations between fields. The most populated region, IC~5070, previously studied in \citet{2021MNRAS.506.5989F, 2023MNRAS.520.5433H, 2024MNRAS.529.4856H}, is over-plotted in Fig.~\ref{Per_hist}. We find 20 (65 percent) fast and 11 (35 percent) slow rotators, consistent with earlier results and indicating a slightly higher fraction of fast rotators than in the full sample. For most other regions, small number statistics limit direct comparisons. However, combining spatially related fields yields more robust estimates (Table~\ref{fast_slow}). We group IC~1396~A and N into ‘IC~1396’, and M~42, L~1641~N, V~898~Ori, YY~Ori, and V~555~Ori into ‘Orion’, while $\sigma$~Ori is treated separately. Despite the diversity of environments, the fast/slow fractions are broadly consistent: Orion and IC~5070 show roughly two-thirds fast and one-third slow rotators, while IC~1396 is close to an even split. Though the statistics are poor for the $\sigma$~Ori region, it stands out with nearly 90 percent of its 14 objects being slow rotators and (as discussed in the next subsection) about 40 percent exhibiting a $K-W2$ excess.

\begin{table}
\caption{\label{fast_slow} Number and fraction of fast / slow rotators and $K-W2$ excess sources in selected HOYS target fields, sorted by increasing fraction of slow rotators. See text for details how the fields are defined. Note that the numbers in the row 'Total' include the sum of sources in the entire HOYS sample.}
\centering
\setlength{\tabcolsep}{5.0pt}
\renewcommand{\arraystretch}{1.0}
\begin{tabular} {|l|c|c|c|c|c|c|}
\hline
Region & \multicolumn{2}{c|}{Fast rotators} & \multicolumn{2}{c|}{Slow Rotators} & \multicolumn{2}{c|}{E($K$-$W2$)} \\ \hline
 & N & [\%] & N & [\%] & N & [\%] \\ \hline
Orion & 30 & 71 & 12 & 29 & 9 & 21 \\
IC~5070 & 20 & 65 & 11 & 35 & 17 & 55 \\
IC~1396 & 19 & 45 & 23 & 55 & 9 & 21 \\
$\sigma$~Ori & 2 & 12 & 15 & 88 & 7 & 41 \\ \hline
Total & 79 & 55 & 65 & 45 & 50 & 35 \\
\hline
\end{tabular}
\end{table}

\subsection{Inner Disc Excess}\label{innerdisc}

The bimodality of YSO rotation periods is commonly attributed to disc braking, where magnetic coupling between the star and its inner disc regulates angular momentum \citep[e.g.][]{2014prpl.conf..433B, 2017A&A...599A..23V, 2025ApJ...979...29B}. We therefore use the $K-W2$ colour to identify the inner dust discs in our sample. Objects with excess emission in this colour ($>0.5$~mag) are interpreted as hosting warm dust close to the star, indicative of a dense inner disc \citep{2012A&A...540A..83T}, or thick disc.

The disc excess fraction for the full sample and the four regions defined previously is summarised in Table~\ref{fast_slow}. Overall, one third of the objects show a $K-W2$ excess. We note that the absence of excess does not imply the absence of an inner disc, as dust depleted or anemic discs \citep[e.g.][]{2020A&A...642A..86T} remain undetected in this colour. The disc fraction varies from $\sim$20 percent in Orion and IC~1396 to 40 percent in $\sigma$~Ori and just over 50 percent in IC~5070, consistent with \citet{2024MNRAS.529.1283F} despite the more stringent variability selection applied here.

\begin{figure}
\centering
\includegraphics[angle=0,width=\columnwidth,alt={Three-panel scatter plot comparing stellar rotation period with variability and colour properties for objects in four young stellar regions IC~5070, $\sigma$~Ori, IC~1396, and Orion. The regions are shown using different coloured points. In the top two panels, the horizontal axis shows rotation period in days, in the bottom panel the horizontal axis shows the colour slope $\alpha_{\mathrm{SED}}$. Dotted guide lines mark key reference values used to separate populations. The upper panel shows rotation period against $K-W2$ colour. Most objects with short periods, below about 5.5~d, have relatively low $K-W2$ values, although a few show stronger infrared excess. At longer periods, the points are more widely spread, and several objects have $K-W2$ values above 0.5~mag, consistent with stronger disc-related infrared excess.}]{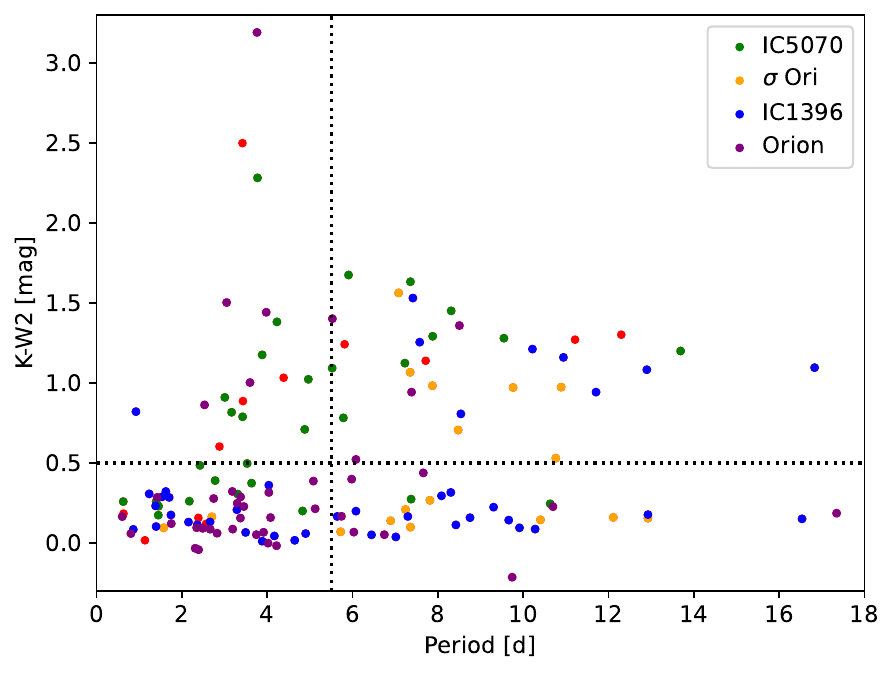} \\\
\includegraphics[angle=0,width=\columnwidth,alt={The middle panel shows rotation period against $\alpha_{\mathrm{SED}}$. The dotted horizontal lines separate YSO classes, with objects below $-2$ corresponding to Class~3 sources, objects between $-2$ and $0$ corresponding to Class~2 sources, and objects above $0$ corresponding to Class~1 sources. The plot shows a mixture of YSO classes across the period range, with several longer-period objects lying in the Class~2 or Class~1 regions.}]{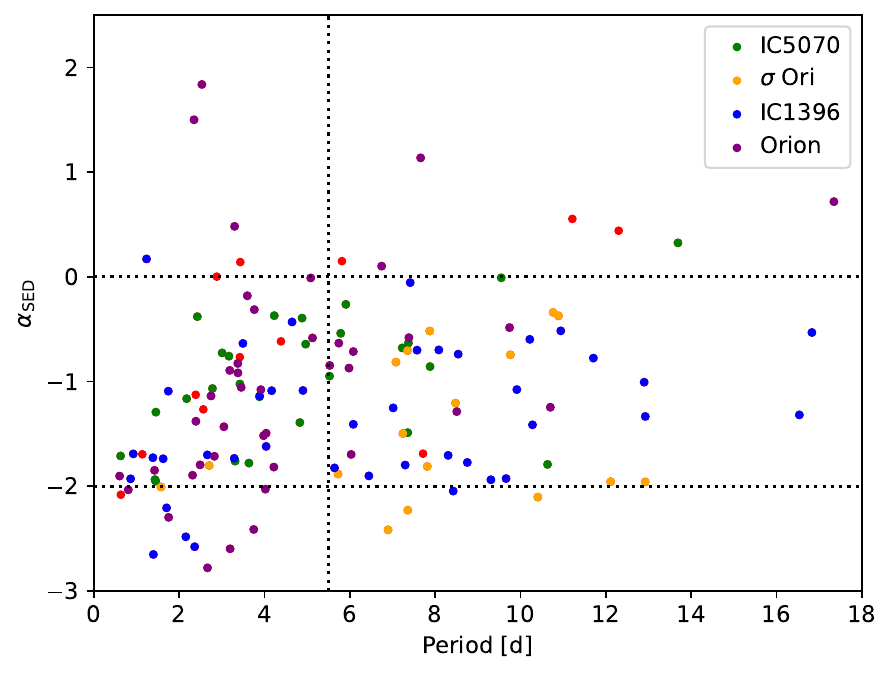} 
\includegraphics[angle=0,width=\columnwidth,alt={The lower panel compares $\alpha_{\mathrm{SED}}$ with $K-W2$ colour. Objects with more negative $\alpha_{\mathrm{SED}}$, characteristic of objects without a disc, generally have small $K-W2$ colours. Objects with higher $\alpha_{\mathrm{SED}}$, indicating stronger disc emission, tend to have larger $K-W2$ values. Overall, the figure illustrates the relationship between rotation period, infrared excess, and spectral energy distribution slope, with disc-bearing objects tending to occupy regions of higher $K-W2$ colour and less negative $\alpha_{\mathrm{SED}}$. The top and bottom panels show some separation between objects with smaller and larger $K-W2$. There is no correlation of properties based on region.
}]{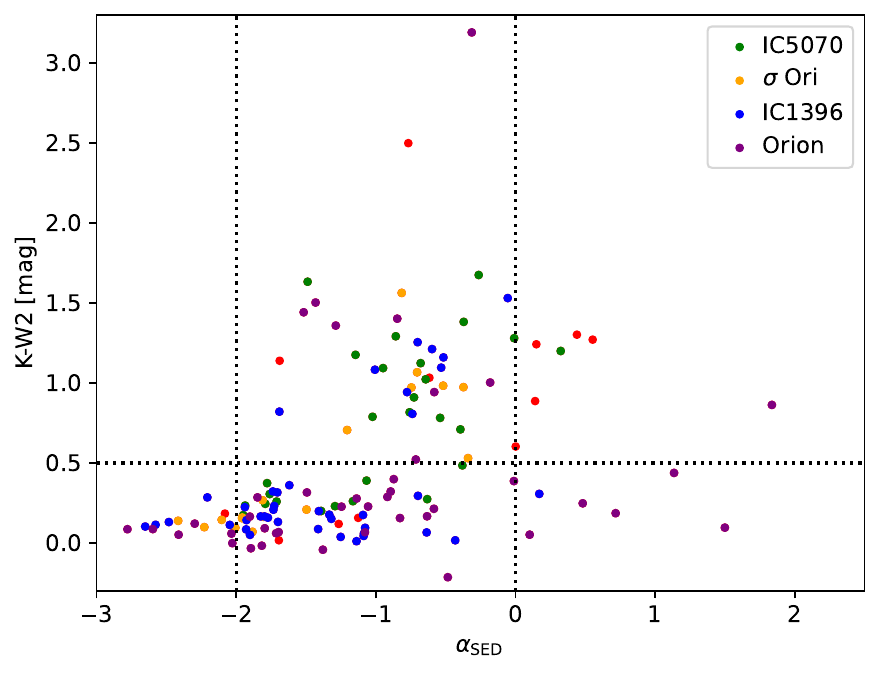} \\ 
\caption{\label{Per_KW2} Relationship between the period, $K-W2$ colour, and $\alpha_{\rm SED}$ slope for the rotational variables in our sample. The colour code indicates the region the sources are in. Red dots indicate objects not in any of the 4 populous regions. Dotted lines separate fast and slow rotators at $P = 5.5$~d, sources with and without colour excess at $K-W2 = 0.5$~mag, and Class~1, 2, and 3 objects based on their $\alpha_{\rm SED}$ value. {\bf Top:} The rotation period vs the $K-W2$ colour. {\bf Middle:} The rotation period vs the $\alpha_{\rm SED}$ slope. {\bf Bottom:} The $K-W2$ colour vs the $\alpha_{\rm SED}$ slope.}
\end{figure}

The relation between rotation period and $K-W2$ colour is shown in the top panel of Fig.~\ref{Per_KW2}, where vertical and horizontal lines at 5.5~d and 0.5~mag separate fast/slow rotators and disc/no-disc objects, respectively. About 40 percent of the sample are fast rotators without disc excess, while the 45 percent of slow rotators split evenly between disc-bearing and disc-free systems. In the standard disc braking picture, stars evolve from slow rotators with discs to fast rotators without discs. However, about one fifth of the sample consists of slow rotators without detectable dust discs, possibly due to gas-rich but dust-depleted discs or tidal locking in binaries. Conversely, roughly 15 percent of objects are fast rotators with disc excess, which may reflect binarity or weaker magnetic coupling. Colour-coding by region shows that slow rotators without disc excess are rare in IC~5070 and are predominantly found in Orion and IC~1396.

The middle and bottom panels of Fig.~\ref{Per_KW2} show the SED slope $\alpha_{\rm SED}$, derived from the $W1$ to $W4$ WISE magnitudes following \citet{2013Ap&SS.344..175M}, as a function of the rotation period and $K-W2$, respectively. The nominal boundaries between Class~1 and Class~2 ($\alpha_{\rm SED}=0$) and between Class~2 and Class~3 ($\alpha_{\rm SED}=-2$) are indicated. The sample is dominated by Class~2 objects, with few Class~1 or Class~3 sources. Within the Class~2 population, objects with disc excess tend to have higher $\alpha_{\rm SED}$ values. A threshold at $\alpha_{\rm SED}=-1$ separates Class~2 objects with and without inner dust discs, which we refer to as early and late Class~2. Accordingly, 70 percent of early Class~2 objects show disc excess, while 85 percent of late Class~2 objects do not. With the exception of very short rotation periods ($P < 2$~d), the $\alpha_{\rm SED}$ slope shows no correlation with the rotation period of the source. This suggests that the innermost parts of the disc are the dominant component regulating angular momentum transfer. Once a small (fraction of an AU) hole is present in the inner disc, the star may no longer be efficiently coupled to the disc, irrespective of how much material remains at larger radii.

\begin{figure*}
\centering
\includegraphics[angle=0,width=0.98\columnwidth,alt={A colour magnitude diagram, absolute $G$ magnitude against $BP-RP$. The objects are colour coded into four groups, fast rotation no disc, slow rotation no disc, fast rotation with disc and slow rotation with disc. A one million year (dashed) and four million year (solid) isochrones are plotted on the figure.}]{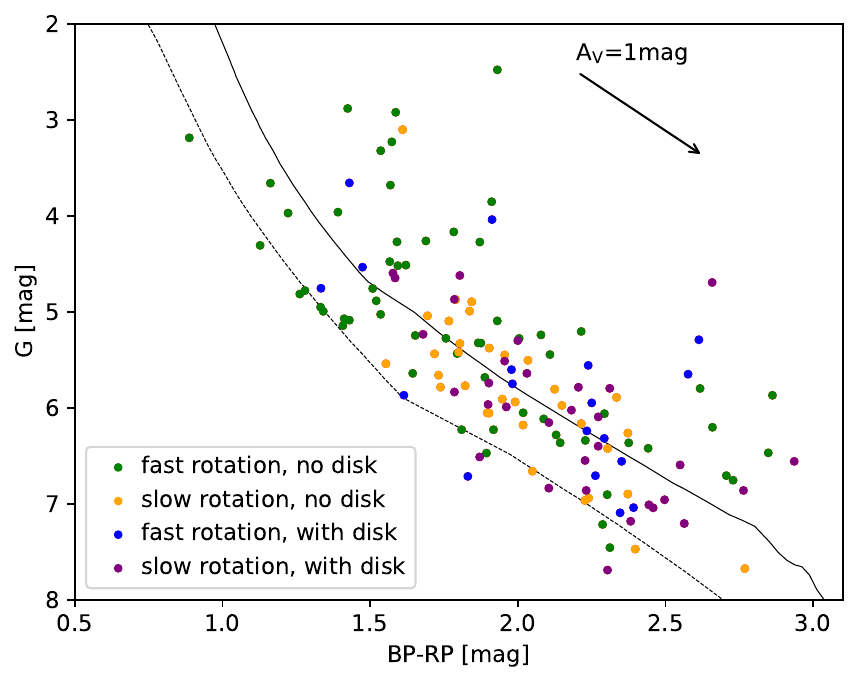} \hfill
\includegraphics[angle=0,width=0.98\columnwidth,alt={The same as left but the objects are colour coded into the star formation region they are in, IC~5070, $\sigma$~Ori, IC~1396 and Orion. The figure shows there is no clear trends based on either groupings.}]{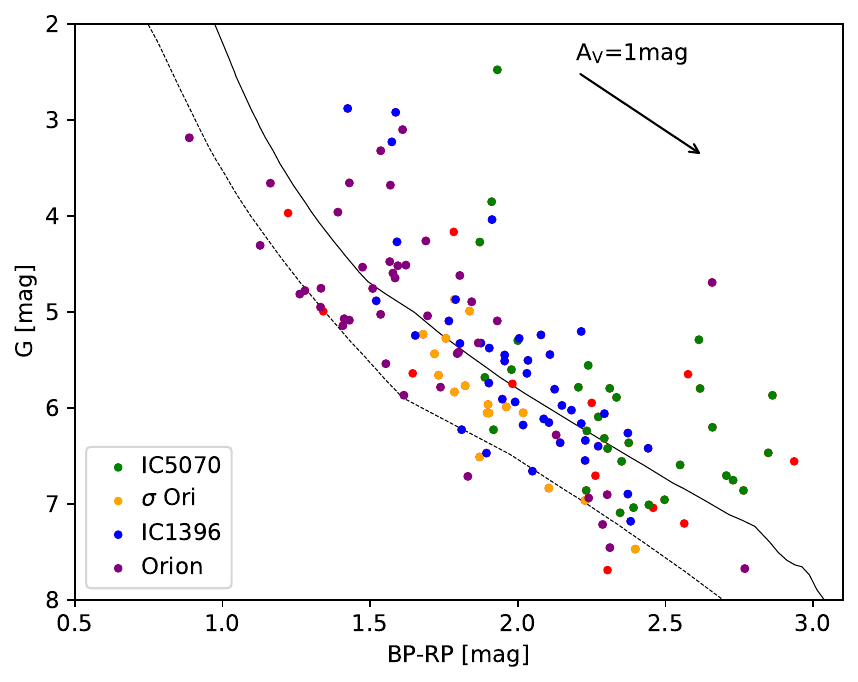}
\caption{\label{CMD_Per_KW2} Gaia absolute magnitude vs. colour of the rotational variables. In the left panel the objects are colour coded based on their distribution in the period vs. $K-W2$ colour plot, and in the right panel according to the star formation region they are in. In all panels the solid black line is a 1~Myr PARSEC isochrone and the dashed black line a 4~Myr isochrone \citep{2012MNRAS.427..127B}. }
\end{figure*}

The Gaia colour–magnitude diagram of the sample is shown in Fig.~\ref{CMD_Per_KW2}, using absolute magnitudes derived from inverted parallaxes without extinction correction. PARSEC isochrones for 1 and 4~Myr \citep{2012MNRAS.427..127B} are over-plotted. Colour coding by period–disc classification (left panel) and region (right panel) reveals no clear trends: objects with and without disc excess are not separated, and neither population is preferentially associated with younger or older isochrones or specific regions. The lower mass limit corresponds to $G \approx 7$~mag, i.e. $\sim$0.6~M$_\odot$ at 4~Myr \citep{2012MNRAS.427..127B}.

Brighter stars ($G<5$~mag, $>1$~M$_\odot$) are predominantly found in Orion, while lower-mass objects are more common in IC~5070. These brighter stars are mainly fast rotators without disc excess; 65 percent fall into this category. This is consistent with more massive stars dispersing their inner discs more rapidly through stronger radiation \citep{2006ApJ...651L..49C}, leading to earlier spin-up.

\subsection{Rotational Variables - Summary}

We have identified a sample of 144 rotationally variable YSOs from 12 different HOYS target fields (young star clusters/star forming regions). Their placement in the colour magnitude diagram suggests that they are mainly younger than 4~Myr, with a median age of  about 1~Myr. Their estimated effective temperatures range from 3500~K to 6500~K,  placing them in an approximate mass range of 0.6 to 2 solar masses. The sample is dominated by Class~2 objects, and one third shows excess emission that can be attributed to warm dust in the inner disc. This supports the standard disc braking scenario in which slow rotators preferentially retain discs while fast rotators are typically disc free. However, substantial populations of slow rotators without dust excess and fast rotators with discs indicate that additional effects such as gas-rich dust-depleted discs, binarity, or weaker magnetic coupling also influence rotational evolution. Two thirds of the sample are fast rotators, predominantly without disc excess emission. The hotter stars (above 5000~K) are generally fast rotators without disc excess emission. There are no discernible age trends with rotation period, disc excess or location. 

\begin{figure*}
\centering
\includegraphics[angle=0,width=0.99\columnwidth,alt={Four-panel comparison of spot property distributions for three samples, labelled s1, s2, and s3. Sample s1 is the original sample from Herbert et al. (2024), sample s2 uses the same amplitudes as Herbert et al. (2024) but with the fitted effective temperatures from this work, and sample s3 represents the IC 5070 results from this work. The left-hand panels show warm-spot properties, while the right-hand panels show cold-spot properties. The upper panels show spot coverage, and the lower panels show the temperature difference between the stellar photosphere and the spot. In each panel, coloured histograms give the number of objects in each bin and there is a CDF plotted for each sample in respective colours. For warm spots, the coverage distributions are strongly concentrated at low spot coverages, mostly below about 0.15, although sample s3 extends to larger coverages of around 0.4 to 0.45. The warm-spot temperature differences are mostly below about 2000~K, with a smaller number of objects extending to higher values. For cold spots, the coverage distributions are broader, with most objects lying between about 0.1 and 0.35. The cold-spot temperature differences are negative, with most objects lying between about −1500 and −500~K. Overall, the three samples show broadly similar distributions, but the IC~5070 sample extends to somewhat larger warm-spot coverages and includes a wide range of cold-spot temperature differences.}]{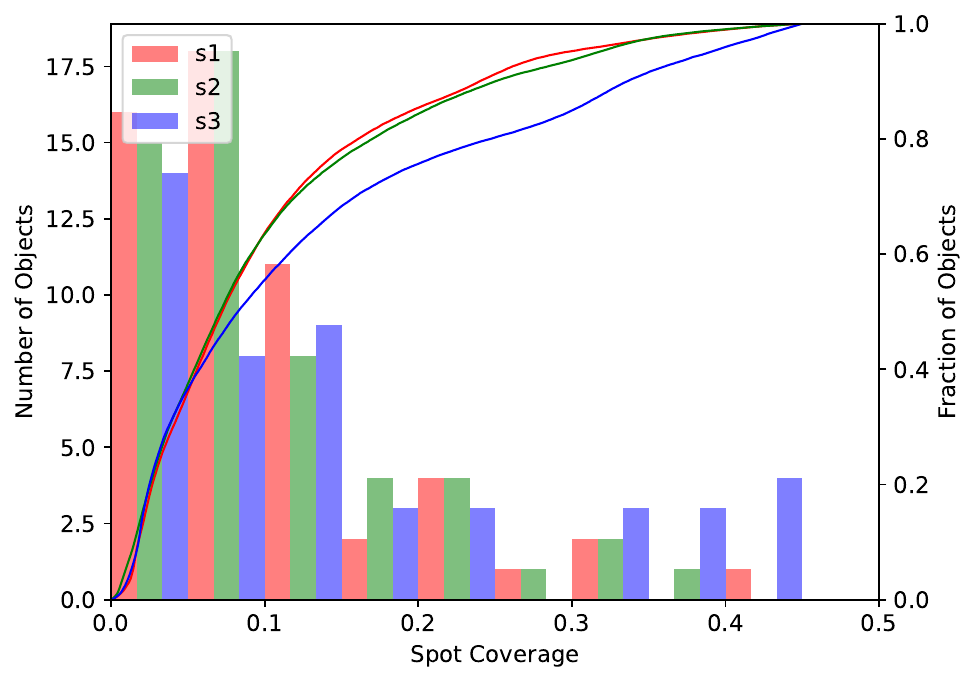} \hfill
\includegraphics[angle=0,width=0.97\columnwidth,alt={Four-panel comparison of spot property distributions for three samples, labelled s1, s2, and s3. Sample s1 is the original sample from Herbert et al. (2024), sample s2 uses the same amplitudes as Herbert et al. (2024) but with the fitted effective temperatures from this work, and sample s3 represents the IC 5070 results from this work. The left-hand panels show warm-spot properties, while the right-hand panels show cold-spot properties. The upper panels show spot coverage, and the lower panels show the temperature difference between the stellar photosphere and the spot. In each panel, coloured histograms give the number of objects in each bin and there is a CDF plotted for each sample in respective colours. For warm spots, the coverage distributions are strongly concentrated at low spot coverages, mostly below about 0.15, although sample s3 extends to larger coverages of around 0.4 to 0.45. The warm-spot temperature differences are mostly below about 2000~K, with a smaller number of objects extending to higher values. For cold spots, the coverage distributions are broader, with most objects lying between about 0.1 and 0.35. The cold-spot temperature differences are negative, with most objects lying between about −1500 and −500~K. Overall, the three samples show broadly similar distributions, but the IC~5070 sample extends to somewhat larger warm-spot coverages and includes a wide range of cold-spot temperature differences.}]{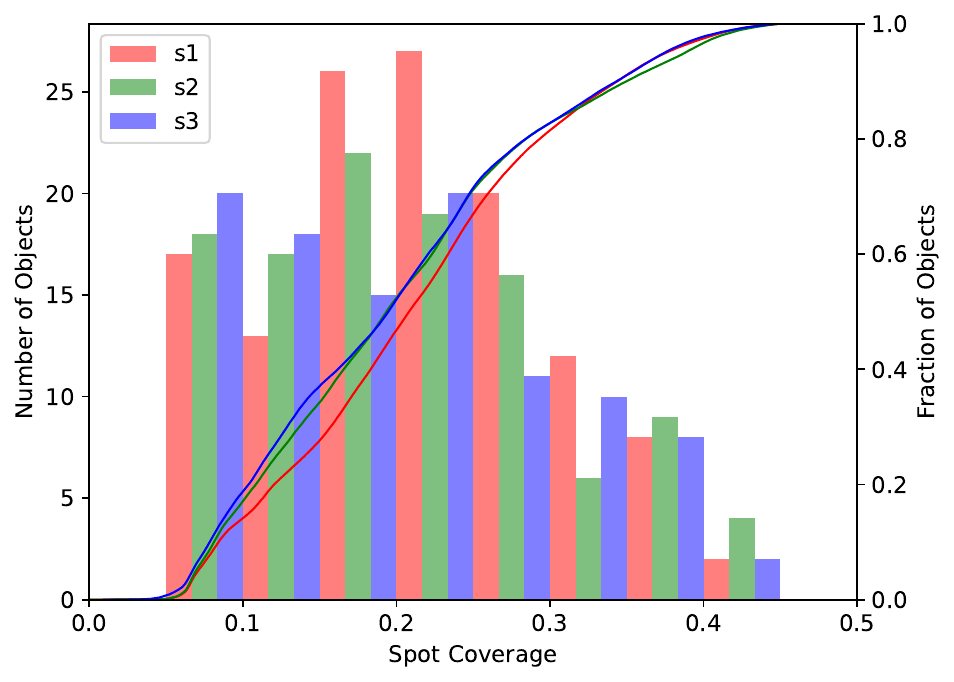} \\ 
\includegraphics[angle=0,width=0.99\columnwidth,alt={Four-panel comparison of spot property distributions for three samples, labelled s1, s2, and s3. Sample s1 is the original sample from Herbert et al. (2024), sample s2 uses the same amplitudes as Herbert et al. (2024) but with the fitted effective temperatures from this work, and sample s3 represents the IC 5070 results from this work. The left-hand panels show warm-spot properties, while the right-hand panels show cold-spot properties. The upper panels show spot coverage, and the lower panels show the temperature difference between the stellar photosphere and the spot. In each panel, coloured histograms give the number of objects in each bin and there is a CDF plotted for each sample in respective colours. For warm spots, the coverage distributions are strongly concentrated at low spot coverages, mostly below about 0.15, although sample s3 extends to larger coverages of around 0.4 to 0.45. The warm-spot temperature differences are mostly below about 2000~K, with a smaller number of objects extending to higher values. For cold spots, the coverage distributions are broader, with most objects lying between about 0.1 and 0.35. The cold-spot temperature differences are negative, with most objects lying between about −1500 and −500~K. Overall, the three samples show broadly similar distributions, but the IC~5070 sample extends to somewhat larger warm-spot coverages and includes a wide range of cold-spot temperature differences.}]{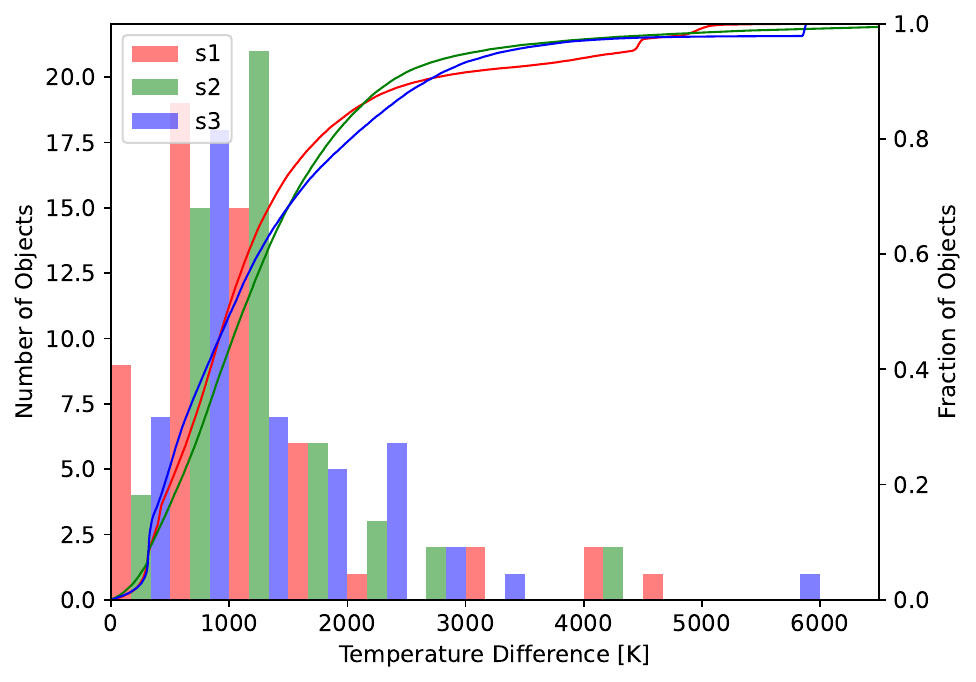} \hfill
\includegraphics[angle=0,width=0.97\columnwidth,alt={Four-panel comparison of spot property distributions for three samples, labelled s1, s2, and s3. Sample s1 is the original sample from Herbert et al. (2024), sample s2 uses the same amplitudes as Herbert et al. (2024) but with the fitted effective temperatures from this work, and sample s3 represents the IC 5070 results from this work. The left-hand panels show warm-spot properties, while the right-hand panels show cold-spot properties. The upper panels show spot coverage, and the lower panels show the temperature difference between the stellar photosphere and the spot. In each panel, coloured histograms give the number of objects in each bin and there is a CDF plotted for each sample in respective colours. For warm spots, the coverage distributions are strongly concentrated at low spot coverages, mostly below about 0.15, although sample s3 extends to larger coverages of around 0.4 to 0.45. The warm-spot temperature differences are mostly below about 2000~K, with a smaller number of objects extending to higher values. For cold spots, the coverage distributions are broader, with most objects lying between about 0.1 and 0.35. The cold-spot temperature differences are negative, with most objects lying between about −1500 and −500~K. Overall, the three samples show broadly similar distributions, but the IC~5070 sample extends to somewhat larger warm-spot coverages and includes a wide range of cold-spot temperature differences.}]{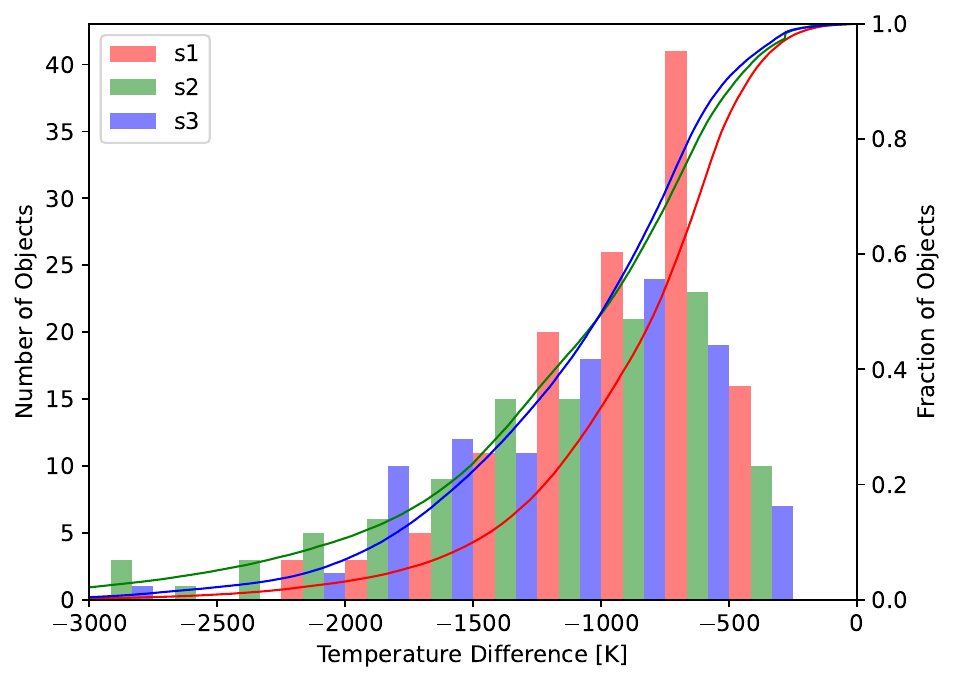} \\ 
\caption{\label{comp_ic5070} Comparison of spot property distributions from \citet{2024MNRAS.529.4856H} and this work for IC~5070. Sample~1 is the original sample from \citet{2024MNRAS.529.4856H}. In sample~2 we used the same amplitudes but our fitted values of \teff. Sample~3 represents this work. Warm spot properties are in the left column, cold spot properties in the right. The top row shows the coverage distribution and the bottom row the star-spot temperature difference distribution.}
\end{figure*}

\section{Characterisation of spot properties}\label{discussion}

\subsection{Method Comparison in IC~5070}

\begin{figure*}
\centering
\includegraphics[width=0.97\columnwidth,alt={A plot of spot temperature difference versus spot coverage for the entire sample of rotational variables. The plot is separated into three sections by two horizontal dashed lines at 0k and 2500k. This separates cold warm and hot spots which are labelled by section. The objects are colour coded according to the HS:CS$_{\{V\}}$ ratio (see Herbert et al. 2024). The figure shows the majority of solutions were cold spots, a small number were hot spot solutions and the remaining solutions warm spots. There is a trend of fewer objects with larger spot coverage consistent across both cold and warm spots with the trend in warm spots being more pronounced.}]{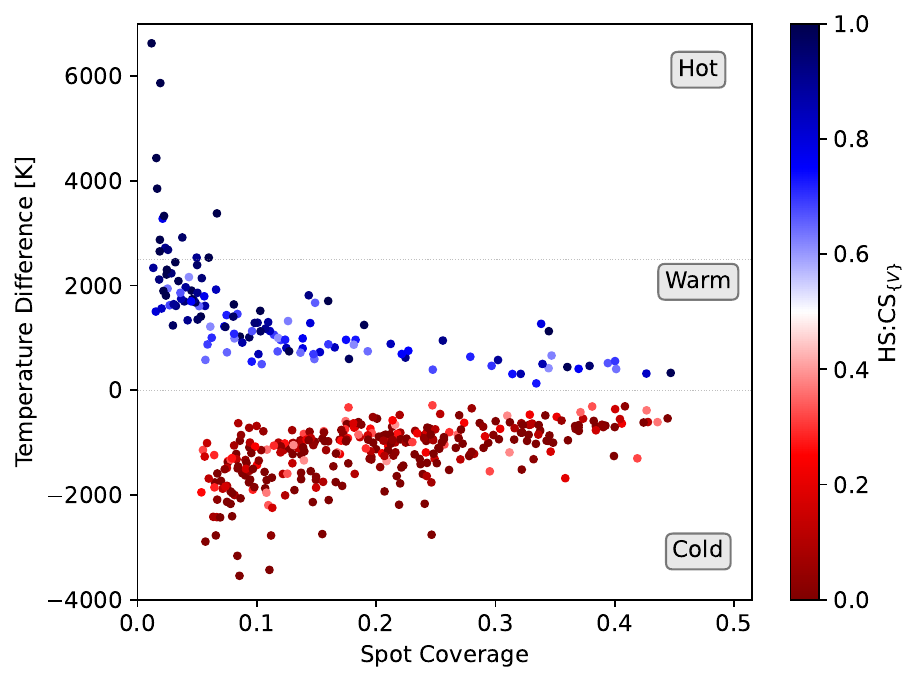} \hfill
\includegraphics[width=1.02\columnwidth,alt={Histogram comparing two spot coverage distributions, shown in red(cold) and blue(hot/warm), with CDFs overplotted using the same colours. The horizontal axis shows spot coverage, ranging from 0 to about 0.45. The left vertical axis gives the number of objects in each bin, while the right vertical axis gives the cumulative fraction of objects. The blue distribution is concentrated at lower spot coverages, with most objects below about 0.15 and only a small number extending to larger coverages. The red distribution is broader and shifted toward higher spot coverages, with many objects between about 0.1 and 0.3 and a tail extending to about 0.45. The cumulative curves show the same behaviour: the blue sample reaches a high cumulative fraction at lower spot coverage, while the red sample increases more gradually, indicating a wider spread of spot coverages.}]{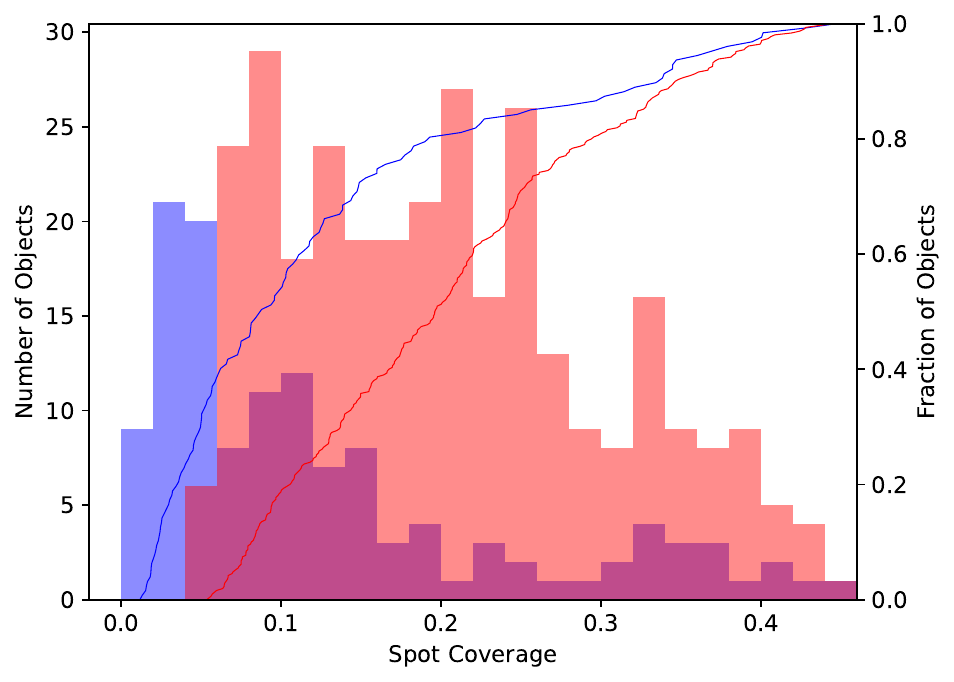}\\
\caption{ {\bf Left:}  Spot temperature difference $T_S - T_\star$ versus spot coverage for the entire sample of rotational variables. Horizontal dotted lines mark $T_S - T_\star = 0$~K and $T_S - T_\star = 2500$\,K, which separate the cold, warm and hot spot solutions. The symbols are colour coded according to the HS:CS$_{\{V\}}$ ratio, our spot property quality indicator. Values close to zero indicate well constrained cold spot properties. Values near one indicate well constrained warm/hot spot properties. Values near 0.5 correspond to increasingly degenerate hot and cold spot solutions \citep[see][for details]{2024MNRAS.529.4856H}. The typical uncertainties of the spot coverage and temperature differences are 0.02 and 250~K, respectively. {\bf Right:} Distribution of spot coverage of hot/warm (blue), and cold spots (red). The solid lines are the CDFs of the hot/warm (blue) and the cold spots (red). \label{fig: spot_prop}}
\end{figure*}

In the following sections, we discuss the distributions of spot properties derived for our sample, based on amplitude measurements of periodic light curves and fitted stellar effective temperatures. We follow the spot fitting procedures outlined in \citet{2023MNRAS.520.5433H,2024MNRAS.529.4856H} with very minor adjustments discussed in Sect.~\ref{spotfitting}. In \citet{2024MNRAS.529.4856H}, the spot properties in IC~5070 were determined using measured \teff\ values from \citet{2020ApJ...904..146F}. Here, we first compare those results with the updated \teff\ distributions obtained in this work.

Fig.~\ref{comp_ic5070} shows histograms and CDFs of spot coverage and spot–star temperature differences ($T_S - T_\star$) for warm (left) and cold (right) spots, derived using three approaches. Sample~1 (red) corresponds to \citet{2024MNRAS.529.4856H}. Sample~2 (green) uses the same amplitudes but updated fitted \teff\ values, while sample~3 (blue) represents the full analysis of this work, including minor amplitude differences caused by small changes in the start dates of light curve slices. The general distributions are consistent, with small systematic offsets mainly due to the systematically higher \teff\ values adopted here.

For cold spots, the temperature difference between the spot and the star increases by $\sim$200~K, approximately half the typical increase in \teff, consistent with the simulations in \citet{2023MNRAS.520.5433H}. The cold spot coverage decreases by $\sim$0.01 when using the fitted \teff. The small differences between samples 2 and 3 indicate that the correspondingly small differences in the measured amplitudes do not significantly affect the derived spot property distributions. Thus, the updated \teff\ values introduce only small systematic shifts, comparable to the typical uncertainties, while providing a more realistic \teff\ distribution (left panel of Fig.~\ref{teff_fit}) and more reliable cold spot properties.

For warm spots, the changes are smaller and less systematic. The temperature difference increases again slightly ($\sim$200~K) when using the fitted \teff-values, while the coverage distribution remains largely unchanged. Variations in amplitude measurements have a somewhat larger effect, producing a slightly higher fraction of large warm spots. However, these are few in number and their properties are less certain (see Sect.~\ref{spot_prop_dist}).

The distributions of spot coverage and temperature difference for the full sample are shown in the left panel of Fig.~\ref{fig: spot_prop}, with histograms and CDFs of spot coverage for cold and warm spots in the right panel. These are discussed in the following subsections.

\begin{figure}
\centering
\includegraphics[width=0.97\columnwidth,alt={A plot of stellar rotation periods versus for all cold spots in the HOYS sample. The vertical axis shows spot coverage from 0.05 to 0.45 and the horizontal axis shows period from 0 to 18~d. There is a sparsely populated region of the plot bound by a black dotted rectangle whose left edge is at a period of approximately 5.5~d and its top side is at spot coverage of 0.20. The figure shows that there is an absence of cold spot solutions with coverage less than 0.20 and a period of greater than six days.}]{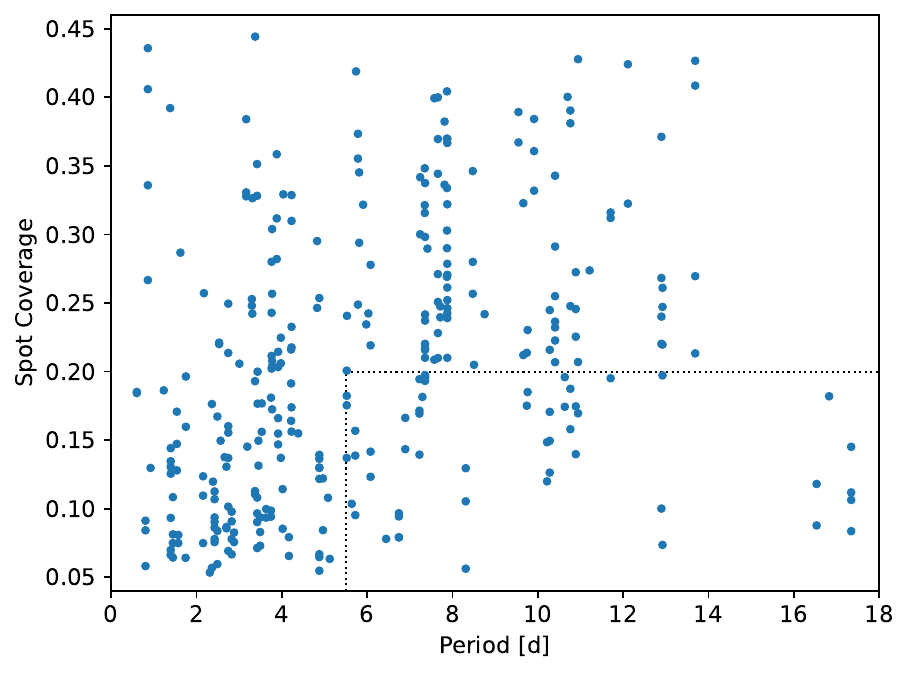}
\caption{ Stellar rotation periods vs. the coverage for all cold spots in our sample. A lack of small spots on slow rotators is evident (bounded by the black dotted lines). The typical uncertainties of the spot coverage are 0.02. The period uncertainties are typically smaller than the symbol size. \label{fig: per_size_col}}
\end{figure}

\subsection{Cold Spot Properties}

The distribution of cold spot coverage (red histogram in the right panel of Fig.~\ref{fig: spot_prop}) is very similar to that found for IC~5070 in \citet{2024MNRAS.529.4856H}, showing an approximately flat distribution between the detection limit (coverage of 0.05) and $\approx 0.25$. To recover the underlying true distribution, this observed distribution must be corrected for the detection efficiency of our method using the simulations described in Sect.~\ref{per_dis} (see also Fig.~\ref{Per_det_rate}). This is addressed in Sect.~\ref{spot_prop_dist}. Before doing so, we assess whether additional selection effects or biases could influence the observed distribution.

The key feature of the data is the dependence of spot coverage on the rotation period (Fig.~\ref{fig: per_size_col}). The fast rotator population ($P < 5.5$~d) contains a much higher fraction of low coverage spots and very few large coverage spots. On the other hand, slow rotators ($P > 5.5$~d) are dominated by large spot coverage ($>0.20$), with comparatively few small spots. This behaviour is not expected from detection biases. The simulations in Fig.~\ref{Per_det_rate} demonstrate that the detection rate is essentially independent of the rotation period. Hence, the observed trend is likely physical, reflecting a combination of shorter lifetimes, non-existence of small spots, and/or smaller temperature contrasts for low-coverage spots on slow rotators. There are two main explanations for why small cold spots could not exist on the slow rotators. The magnetic pressure on the spatial scale of the spots is lower than the gas pressure in the surrounding stellar atmosphere. The granulation scale on the stars is larger than the spot size. Both are very unlikely given the nature of the YSOs and the coverage range (0.05 to 0.20) of the 'missing' cold spots. 

We briefly consider and exclude other possible explanations. The simulations rule out a detection bias against low-coverage spots at longer periods. A bias in the effective temperature is also unlikely. When we restrict the sample to \teff\ $< 5500$~K (where both fast and slow rotators are present), it does not alter the observed distribution. Thus, the \teff\ distribution cannot account for the trend. Consistent with studies of main-sequence stars, where larger spots and spots on cooler stars have longer lifetimes \citep{2017MNRAS.472.1618G}, the most plausible explanations are that low-coverage spots on slow rotators either have reduced temperature contrast, or have lifetimes too short to be detected by our methodology. These will be briefly discussed in the next two subsections. Furthermore, using homogeneous period and coverage distributions, a simple Monte-Carlo simulation shows that there is less than $6 \times 10^{-5}$ chance that the observed distribution is a statistical fluke.

\subsubsection{Cold Spot Temperature Contrast}

The key result is that variations in the spot–star temperature contrast cannot explain the lack of low-coverage spots on slow rotators. As shown in the left panel of Fig.~\ref{fig: spot_prop}, there is a weak trend in the cold spot properties: spots with large temperature differences ($> 2000$~K) predominantly have small coverage. This implies that smaller spots tend to exhibit higher temperature contrasts than larger cold regions. We note that the derived spot temperatures represent averages over the full spot area, and localised regions within the spots may reach lower temperatures.

There is a detection bias, as the sensitivity to spot–star temperature differences increases with spot coverage. For a given temperature contrast, larger spots produce higher amplitudes and are therefore easier to detect. To account for this, we impose a common lower limit on the temperature difference (e.g. 750~K). This slightly modifies the period-coverage distribution, but does not remove the main trend. In particular, the deficit of low-coverage spots among slow rotators persists, as the selection primarily removes larger spots across all periods. Hence, while reduced temperature contrasts contribute to the observed behaviour in Fig.~\ref{fig: per_size_col}, they are not the dominant effect.

\subsubsection{Cold Spot Lifetimes}

Our period determination (Sect.~\ref{per_amp}) is based on 6~month light-curve segments. As a result, only spots that remain stable for a significant fraction of this timescale can be detected. In practice, this implies that only spots with lifetimes $\gtrsim$150–200~d are included in our analysis. The absence of small spots on slow rotators (Fig.~\ref{fig: per_size_col}) therefore directly indicates that such spots must have lifetimes shorter than this threshold and are therefore not recovered by our method. In combination with the lack of biases discussed above, this strongly suggests that spot lifetime is the primary driver of the observed distribution.

This interpretation can be placed in the framework of turbulent magnetic diffusivity. \citet{2014ApJ...795...79B} showed that the empirical spot lifetime–size relation can be explained if the relevant diffusion scale is set by the supergranule size, with only a weak dependence on other stellar parameters. Using their Fig.~1, we can estimate whether our results are consistent with this picture. For fast rotators, the detection limit of a coverage of 0.05 corresponds to spot radii of about 1.5–$2.2 \times 10^5$~km (assuming single spots on stars with radii of 1–$1.5 R_\odot$). These spots persist for at least 150–200~d, which implies supergranule sizes of $\lesssim 0.1$ solar radii. In contrast, for slow rotators, spots with coverage $<0.2$ (radii about 3.0–$4.5 \times 10^5$~km) are rarely detected, indicating lifetimes shorter than six months. Within the same framework, this implies substantially larger supergranules, of the order of half the solar radius or more.

These results suggest that spot lifetimes, and hence supergranule sizes, depend on stellar rotation. Faster rotating stars appear to host smaller supergranules, allowing spots of a given size to survive longer. This is consistent with the established proportionality between spot size and lifetime for the Sun and main-sequence stars \citep{1938IzPul..16...36G, 1955epds.book.....W, 1997SoPh..176..249P, 2005LRSP....2....8B}, and likely extends to YSOs \citep[see also][]{2017MNRAS.472.1618G}. We note that our measurements represent asymmetrically distributed spot coverage rather than individual spot sizes, and thus the inferred radii are upper limits.

In summary, the lack of small spots on slow rotators is best explained by short spot lifetimes below $\sim$6~months. Within the turbulent diffusivity framework, this implies that supergranule sizes decrease during the spin-up phase after disc dispersal, potentially by at least a factor of five. The physical origin of supergranule sizes remains uncertain even for the Sun \citep{2014ApJ...795...79B}, but the apparent dependence on rotation suggests a link to changes in magnetic field structure during early stellar evolution. A more detailed analysis of spot lifetimes as a function of coverage and rotation is required to quantify this effect, but this is beyond the scope of this work.

\subsubsection{Cold Spot Coverage Distribution}\label{spot_prop_dist}

\begin{figure}
\centering
\includegraphics[width=0.97\columnwidth,alt={Histogram showing the detection efficiency corrected spot coverage distribution for cold spots on fast rotators. The horizontal axis shows spot coverage, ranging from 0 to about 0.45, and the vertical axis shows the fraction of objects. The red histogram peaks at low spot coverages, around 0.08 to 0.10, and then declines toward larger coverages. A small number of objects extend to higher spot coverages of about 0.3 to 0.45. A blue exponential fit is over plotted on the histogram. The fitted curve rises sharply at low spot coverage, peaks near 0.08, and then decreases steadily as spot coverage increases. Overall, the figure shows that cold spots on fast rotators are most commonly found at relatively low covering fractions, while larger spot coverages are less frequent.}]{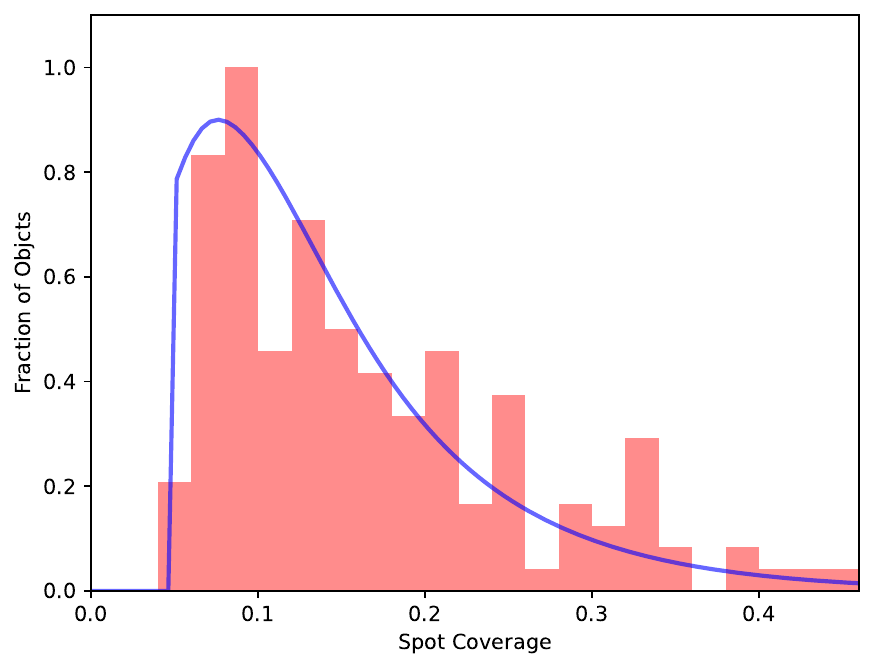}
\caption{Exponential fit to the coverage distribution of cold spots on fast rotators. The fit has been corrected for the detection efficiency shown in Fig.~\ref{Per_det_rate}. \label{fig: cov_fit_col}}
\end{figure}

In Fig.~\ref{fig: spot_prop} we present the distribution of spot properties for the full sample, extending Fig.~5 of \citet{2024MNRAS.529.4856H} to a sample $\sim$2.5 times larger. As shown in the previous subsection, the spot coverage distribution depends on rotation period, with only fast rotators ($P < 5.5$~d) hosting spots whose lifetimes exceed the 6~month detection limit across the full coverage range. We therefore restrict the analysis of the intrinsic coverage distribution to these objects (Fig.~\ref{fig: cov_fit_col}).

The observed distribution peaks at a coverage just below 0.1, and at a star--spot temperature difference near 1000~K (not shown). After correcting for detection efficiency using the simulations described in Sect.~\ref{per_dis} (see also Fig.~\ref{Per_det_rate}), we test different functional forms for the underlying spot coverage distribution. Log-normal and power-law models fail to reproduce the data, while an exponential distribution provides a good fit (solid blue line in Fig.~\ref{fig: cov_fit_col}). This demonstrates that small spot coverages are intrinsically much more common than large ones, with no preferred coverage beyond a characteristic scale. Such behaviour is naturally explained if spot formation is governed by stochastic magnetic flux emergence rather than by a process that produces a characteristic spot size. The temperature contrast distribution cannot be modelled in the same way because the rise towards the peak at $\sim$1000~K is strongly affected by selection effects. However, beyond this peak the number of spots decreases rapidly with increasing temperature contrast, qualitatively consistent with stronger suppression of convection being progressively less common.

For slow rotators ($P > 5.5$~d), the observed distribution peaks at a coverage of $\sim$0.2 and a temperature difference just below 1000~K, consistent with the lifetime-driven detection bias discussed above. However, the number statistics are insufficient to constrain the underlying distribution for this subsample.

\subsection{Hot/Warm Spot Properties}

The coverage distribution of hot/warm spots indicates the presence of two distinct populations (right panel in Fig.~\ref{fig: spot_prop}). The distribution is bimodal, with peaks near 0.05 and 0.1, implying two approximately equally populated groups: small hot spots consistent with accretion column footprints, and larger warm spots likely associated with plages or faculae.

There are significantly fewer hot/warm spots (128) than cold spots (311) in our sample. In addition, some solutions for large coverage, warm spots are less reliable, as indicated by low HS:CS$_{{V}}$ values (see \citealt{2024MNRAS.529.4856H} for details). Consequently, the following discussion remains qualitative. \citet{2024MNRAS.529.4856H} reported an apparent gap between the hot and the warm spots in the spot–star temperature difference distribution. With the improved sample size, this gap is no longer present. Instead, the temperature distribution is smooth. Unlike for cold spots, there is no clear deficit of small warm/hot spots on slow rotators, although this may be influenced by limited number statistics.

The larger warm spots identified are consistent with magnetic surface features such as plages or faculae \citep[see also][]{2020MNRAS.497.4602F}. If this interpretation is correct, these structures should have a latitude distribution similar to that of the cold spots. However, confirming this would require detailed modelling of phase-folded light curves, which is beyond the scope of this work.

\section{Conclusion}

We conducted the first homogeneous, multi-region analysis of rotational variability and spot properties in YSOs using long-term, multi-band photometry from the HOYS project. From an initial sample of over 2000 candidate members across 25 star-forming regions, we identified 144 YSOs with robust periodic signals and well-constrained amplitudes in three optical bands. These objects span 12 regions and occupy a well-defined locus in colour–magnitude space consistent with ages younger than $\sim$4~Myr, with a typical age of the order of 1~Myr. Their effective temperatures range from approximately 3500~K to 6500~K, corresponding to a mass range of roughly 0.6~–~2~M$_\odot$. The sample is dominated by Class~2 objects, with about one third showing clear evidence of warm dust in the inner disc.

The distribution of rotation periods is strongly bimodal, with 55 percent of the objects classified as fast rotators ($P < 5.5$~d) and 45 percent as slow rotators. This bimodality is robust against observational biases and is consistently recovered across multiple regions, in agreement with previous work. The relationship between rotation and inner disc indicators supports the disc braking paradigm: fast rotators are predominantly disc-less, while slow rotators comprise both disc-bearing and disc-free systems. At the same time, the presence of slow rotators without detectable dust discs and fast rotators with disc signatures demonstrates that disc braking alone cannot fully account for the observed rotational states, highlighting the role of additional physical processes such as magnetic field strength, binarity, or rapid disc evolution.

By combining multi-band amplitudes with photometrically derived effective temperatures, we have characterised the statistical properties of stellar spots across this diverse sample. For fast rotators, the intrinsic cold spot coverage distribution is well described by an exponential function once observational biases are removed. This implies that small spot coverages are intrinsically much more common than large ones and suggests that spot formation is governed by stochastic magnetic flux emergence rather than by a process producing a preferred spot size. This constitutes the first empirical constraint on the intrinsic spot coverage distribution of YSOs. In contrast, slow rotators show a pronounced deficit of small-coverage spots. Given that our simulations demonstrate a negligible dependence of detectability on rotation period, this difference is intrinsic and points to a fundamental change in spot evolution between the populations of fast and slow rotators. The inferred spot property distributions are robust against plausible systematic uncertainties in the adopted stellar effective temperatures and other methodological choices discussed in Sect.~\ref{spotfitting}, indicating that the conclusions presented here do not depend on the precise temperature calibration.

The most significant result of this work is that spot evolution differs fundamentally between fast and slow rotators. The absence of small, long-lived spots on slow rotators suggests that these features evolve or decay on timescales shorter than the six-month window required for detection, while comparable spots on fast rotators remain stable. Within the framework of turbulent magnetic diffusion, this implies that the characteristic scales of surface convection, and therefore magnetic structuring, differ significantly between fast and slow rotators. In other words, our data suggest that the supergranule sizes in YSOs are influenced by the rotation period. Although this interpretation relies on indirect inference, the consistency of the observed trends across the sample provides compelling evidence for a rotation-dependent regime of magnetic surface structure in young stars.

Warm and hot spot solutions reveal a more complex picture, with indications of two distinct populations: small, high-temperature contrast features consistent with accretion funnel footprints, and larger, moderate-temperature structures that may represent plage- or faculae-like activity. Although the number statistics are limited, these results suggest that both magnetic activity and accretion processes contribute to the observed rotational variability in YSOs.

Together, our results establish an observational picture in which rotation, disc evolution, and magnetic surface structure are closely linked in young stars. The combination of a bimodal rotation distribution, a partially, but not exclusively, disc-regulated spin evolution, and a rotation-dependent cold spot coverage distribution provides new empirical constraints on models of early stellar angular momentum evolution and magnetic activity. Future work incorporating improved stellar parameter estimates, extinction corrections, and time-resolved spot modelling will be essential to refine these conclusions and to fully exploit the unique long-term monitoring capabilities of the HOYS survey.

\section*{Acknowledgements}

We would like to thank all contributors of observational data for their efforts towards the success of the Hunting Outbursting Young Stars project. CH is supported by the Science and Technology Facilities Council under grant number STFC Kent 2020 DTP ST/V50676X/1. HS is supported by the Science and Technology Facilities Council under grant number STFC Kent ST/Y509267/1.


\section*{Data Availability Statement}

The data underlying this article are available in the HOYS database at https://astro.kent.ac.uk/HOYS-CAPS/.


\bibliographystyle{mnras}
\bibliography{bibliography} 


\appendix

\onecolumn
\section{YSO Rotation Periods}

\begin{longtable}{|c|c|c|c|c|c|}
\caption{\label{rotationperiods} Periodic YSOs identified in this work. We list the Gaia DR3 ID number, the RA and DEC of the source, as well as the period and it's uncertainty measured in our data, and the adopted effective temperature. See text for details on the temperature, period and uncertainty determination.} \\

\hline
Gaia DR3 ID & RA & DEC & Period & Period Uncertainty & \teff \\
\hline
 & \multicolumn{2}{c|}{(J2000) [deg]} & \multicolumn{1}{c|}{[days]} & \multicolumn{1}{c|}{[days]}& \multicolumn{1}{c|}{[K]} \\
\hline
\endfirsthead

\hline
Gaia DR3 ID & RA & DEC & Period & Period Uncertainty & \teff \\
\hline
 & \multicolumn{2}{c|}{(J2000) [deg]} & \multicolumn{1}{c|}{[days]} & \multicolumn{1}{c|}{[days]} & \multicolumn{1}{c|}{[K]} \\
\hline
\endhead

\hline
\multicolumn{6}{|r|}{\textit{Continued on next page}} \\
\hline
\endfoot

\hline
\endlastfoot
1974542739683523584 & 328.2868 & 47.11833 & 2.56542 & 0.00022 & 4790 \\
1974545282304297856 & 328.3771 & 47.24553 & 5.81396 & 0.00031 & 3911 \\
1974545625901682816 & 328.4204 & 47.26174 & 12.299 & 0.043 & 3599 \\
1974545694621155584 & 328.4209 & 47.28096 & 3.42041 & 0.00064 & 3848 \\
1974546072578275584 & 328.3576 & 47.26427 & 11.215 & 0.034 & 4016 \\
1974546450535495680 & 328.4107 & 47.28600 & 2.88 & 0.11 & 4761 \\
1974731890034857600 & 328.1940 & 47.21388 & 3.43143 & 0.00012 & 4324 \\
1974731993119999616 & 328.1718 & 47.21445 & 7.7144 & 0.0016 & 3721 \\
1974738418391182592 & 328.2666 & 47.39185 & 4.385 & 0.023 & 4333 \\
2066866222797779072 & 312.8774 & 44.06249 & 3.165669 & 0.000034 & 4681 \\
2066866287224242432 & 312.8706 & 44.07308 & 4.82797 & 0.00020 & 6544 \\
2066870689567848192 & 312.7814 & 44.17626 & 3.529858 & 0.000096 & 4526 \\
2067058740416252416 & 312.7431 & 44.24230 & 4.87958 & 0.00029 & 4654 \\
2067058740416252544 & 312.7434 & 44.24562 & 3.42337 & 0.00030 & 4811 \\
2067061042518587264 & 312.4549 & 44.17951 & 3.30949 & 0.00014 & 4303 \\
2162929419847388544 & 313.0956 & 44.01580 & 3.63388 & 0.00057 & 4722 \\
2162934986124720896 & 313.3574 & 44.17924 & 1.45322 & 0.00013 & 4246 \\
2162941273956895872 & 313.3527 & 44.34278 & 2.77937 & 0.00014 & 4226 \\
2162944916089431168 & 313.0939 & 44.23338 & 3.00580 & 0.00027 & 4259 \\
2162947596149035648 & 312.8260 & 44.21893 & 7.87857 & 0.00078 & 4437 \\
2162950413647417728 & 312.9439 & 44.37256 & 2.17499 & 0.00013 & 4378 \\
2162950546789495552 & 313.1453 & 44.23346 & 10.63256 & 0.00027 & 4373 \\
2162952475231874688 & 313.1180 & 44.35405 & 5.7851 & 0.0019 & 4145 \\
2162960481050791936 & 313.4677 & 44.28486 & 7.3719 & 0.0071 & 4623 \\
2162964088823157632 & 313.4243 & 44.36014 & 1.4476257 & 0.0000039 & 5721 \\
2162965252757133056 & 313.3618 & 44.42429 & 0.626 & 0.018 & 5521 \\
2163135573981345280 & 312.7561 & 44.25956 & 5.523 & 0.048 & 3929 \\
2163135578281583744 & 312.7565 & 44.26166 & 7.35702 & 0.00055 & 4524 \\
2163136059317926016 & 312.8131 & 44.30488 & 4.22803 & 0.00040 & 4000 \\
2163136368555566848 & 312.7190 & 44.27889 & 8.312 & 0.089 & 4781 \\
2163136402915307136 & 312.7446 & 44.29188 & 7.2271 & 0.0027 & 4389 \\
2163137261908777472 & 312.7776 & 44.36131 & 1.3974285 & 0.0000019 & 4413 \\
2163137742945115136 & 312.8445 & 44.35210 & 3.882 & 0.021 & 4418 \\
2163138601938577024 & 312.8188 & 44.38277 & 2.42438 & 0.00014 & 4749 \\
2163139594070911360 & 312.6920 & 44.31941 & 13.69 & 0.26 & 3981 \\
2163140319925497088 & 312.7639 & 44.40429 & 5.906 & 0.011 & 3842 \\
2163144271295324544 & 312.9371 & 44.43861 & 3.77296 & 0.00017 & 4064 \\
2163146779556221952 & 313.1092 & 44.57394 & 1.4325902 & 0.0000039 & 4665 \\
2163148772421081728 & 312.8450 & 44.56183 & 4.96431 & 0.00093 & 3828 \\
2163156056685634944 & 312.7258 & 44.63561 & 9.5473 & 0.0031 & 4011 \\
2178374878170801792 & 324.5560 & 57.18633 & 4.9012 & 0.0062 & 4389 \\
2178377901827757056 & 324.6805 & 57.30997 & 10.28020 & 0.00084 & 4653 \\
2178383601237749632 & 324.8094 & 57.37025 & 4.03718 & 0.00041 & 4267 \\
2178386045086501760 & 325.0386 & 57.46087 & 3.29054 & 0.00017 & 4368 \\
2178386938439659392 & 324.9366 & 57.51107 & 0.924 & 0.035 & 5538 \\
2178387556914944896 & 325.0083 & 57.56533 & 8.422 & 0.022 & 4954 \\
2178391199046443648 & 324.4901 & 57.37988 & 7.57471 & 0.00039 & 4597 \\
2178392740924572800 & 324.4370 & 57.40375 & 1.62422 & 0.00014 & 4326 \\
2178392848313851392 & 324.3684 & 57.39025 & 16.53 & 0.36 & 4589 \\
2178393982185180288 & 324.3181 & 57.44453 & 7.414 & 0.011 & 3742 \\
2178394016544915328 & 324.2946 & 57.43619 & 1.3955 & 0.0039 & 6417 \\
2178394291422867840 & 324.4593 & 57.43017 & 2.66041 & 0.00028 & 4414 \\
2178394394502330496 & 324.5275 & 57.45599 & 1.70361 & 0.00013 & 4898 \\
2178394497581292800 & 324.4461 & 57.43459 & 6.0797 & 0.0018 & 5218 \\
2178394531941271936 & 324.4657 & 57.44790 & 2.362 & 0.087 & 6470 \\
2178394699429915904 & 324.4628 & 57.46391 & 8.5386 & 0.0015 & 4066 \\
2178395489703989504 & 324.4816 & 57.49225 & 7.2963 & 0.0017 & 4417 \\
2178395631452868224 & 324.4119 & 57.49360 & 10.217 & 0.039 & 4137 \\
2178395975050257024 & 324.4922 & 57.52218 & 1.3834638 & 0.0000030 & 4495 \\
2178398895628585984 & 324.8588 & 57.52984 & 2.155 & 0.036 & 6198 \\
2178399101787006464 & 324.8732 & 57.56156 & 0.861421747 & 0.000000069 & 4876 \\
2178400098219387904 & 324.8006 & 57.60453 & 16.8292 & 0.0057 & 4647 \\
2178401575688130688 & 324.7310 & 57.59160 & 8.3030 & 0.0040 & 4817 \\
2178402056712608896 & 324.6082 & 57.56924 & 9.30624 & 0.00048 & 4507 \\
2178402400321563904 & 324.5666 & 57.61572 & 8.7529 & 0.0034 & 4862 \\
2178402606480266112 & 324.6897 & 57.63335 & 3.8835 & 0.0093 & 4246 \\
2178402675199735424 & 324.6682 & 57.64373 & 9.9132 & 0.0011 & 4442 \\
2178417346809283456 & 324.0296 & 57.37860 & 1.745 & 0.034 & 4938 \\
2178421195099930240 & 324.0530 & 57.47680 & 4.645 & 0.012 & 5254 \\
2178434938995215232 & 323.8276 & 57.56918 & 11.7067 & 0.0018 & 4438 \\
2178440814510995456 & 324.2248 & 57.46629 & 4.16694 & 0.00050 & 5558 \\
2178440951949966848 & 324.2793 & 57.45021 & 12.8977 & 0.0019 & 4620 \\
2178441261187598208 & 324.2892 & 57.49477 & 3.493 & 0.021 & 5008 \\
2178441432986447744 & 324.2939 & 57.52010 & 10.9430 & 0.0020 & 4308 \\
2178442188900684032 & 324.2753 & 57.53378 & 1.23416 & 0.00047 & 4895 \\
2178443185333123584 & 324.4591 & 57.56120 & 12.924 & 0.061 & 4450 \\
2178446793105656704 & 324.0795 & 57.63622 & 7.01448 & 0.00076 & 4178 \\
2178449438799259520 & 324.5703 & 57.65732 & 8.0844 & 0.0019 & 3699 \\
2178449919842118144 & 324.6450 & 57.68859 & 1.5396700 & 0.0000031 & 4755 \\
2178451019353727104 & 324.5940 & 57.73136 & 6.4447 & 0.0018 & 5109 \\
2178452668621309568 & 324.6846 & 57.79075 & 9.66039 & 0.00097 & 4829 \\
2178498882466387072 & 324.9171 & 57.78227 & 5.6398 & 0.0012 & 4646 \\
2221015863228441856 & 325.6784 & 66.20783 & 0.63188429 & 0.00000050 & 5570 \\
3016871585188001536 & 84.1696 & -6.86793 & 0.60243683 & 0.00000015 & 3826 \\
3016879556647282816 & 84.4641 & -6.66429 & 0.80573197 & 0.00000044 & 5067 \\
3016882374145831680 & 84.4283 & -6.67111 & 3.74654 & 0.00097 & 5085 \\
3016904192579682176 & 84.6331 & -6.60688 & 1.422 & 0.060 & 3697 \\
3016935563020738688 & 84.2457 & -6.48469 & 10.698 & 0.020 & 3618 \\
3016957278375450752 & 84.5426 & -6.42406 & 3.37636 & 0.00060 & 3970 \\
3017197830905390592 & 83.8739 & -5.81251 & 1.7518139 & 0.0000088 & 4859 \\
3017199170935143680 & 84.1010 & -5.74686 & 5.9794 & 0.0065 & 4475 \\
3017237993141611776 & 83.6677 & -5.83767 & 2.3913 & 0.0082 & 5639 \\
3017244216552060672 & 83.6985 & -5.73077 & 4.2183 & 0.0059 & 5324 \\
3017245831459893760 & 83.7997 & -5.76055 & 4.0381 & 0.0018 & 4354 \\
3017246651795678720 & 83.8245 & -5.70969 & 5.52737 & 0.00075 & 5298 \\
3017250087773385344 & 83.8042 & -5.65970 & 3.048676 & 0.000059 & 5886 \\
3017250126427057024 & 83.7124 & -5.70597 & 3.1840 & 0.0065 & 5261 \\
3017251947493281536 & 83.9301 & -5.60763 & 2.6630 & 0.0025 & 5522 \\
3017265377852811776 & 83.8067 & -5.51602 & 8.505 & 0.013 & 5316 \\
3017265760107995904 & 83.7353 & -5.52672 & 5.740282 & 0.000063 & 5227 \\
3017266168126793984 & 83.7625 & -5.50042 & 5.0828 & 0.0044 & 5059 \\
3017266206784582144 & 83.7877 & -5.49979 & 4.07794 & 0.00047 & 5258 \\
3017267306296233216 & 83.6465 & -5.53622 & 5.12 & 0.14 & 5438 \\
3017267478094926720 & 83.6154 & -5.53204 & 3.37257 & 0.00040 & 5957 \\
3017270669252519680 & 83.6883 & -5.41780 & 3.45359 & 0.00024 & 5576 \\
3017291942240221056 & 84.2538 & -5.69361 & 3.1916 & 0.0073 & 5109 \\
3017346952181228416 & 83.8743 & -5.54819 & 3.9141 & 0.0034 & 5795 \\
3017347291470793216 & 83.8849 & -5.51979 & 2.313 & 0.033 & 6301 \\
3017355400367313024 & 84.1339 & -5.39586 & 4.020 & 0.011 & 6516 \\
3017359527845298048 & 83.8955 & -5.48734 & 2.532 & 0.075 & 4294 \\
3017364544367271936 & 83.7113 & -5.40034 & 2.35 & 0.11 & 5190 \\
3017364613086735360 & 83.7332 & -5.38696 & 17.35 & 0.34 & 6176 \\
3017367494996951424 & 83.8269 & -5.27713 & 3.76 & 0.17 & 4345 \\
3017368014699258240 & 84.0437 & -5.32917 & 6.03019 & 0.00042 & 4796 \\
3017371588112037248 & 84.0708 & -5.19522 & 7.38 & 0.32 & 5038 \\
3017373924574255488 & 83.9282 & -5.19855 & 3.29813 & 0.00053 & 4819 \\
3017374233811898880 & 83.9422 & -5.18107 & 3.59665 & 0.00014 & 3976 \\
3017375951801904256 & 83.9652 & -5.13590 & 2.82524 & 0.00065 & 5424 \\
3017382407121669504 & 84.2156 & -5.22499 & 2.48995 & 0.00053 & 5193 \\
3046357806990068096 & 107.0442 & -10.76915 & 1.1351 & 0.0022 & 5861 \\
3046581110929906560 & 107.1821 & -10.41369 & 2.3871 & 0.0061 & 5140 \\
3209521037582290304 & 83.7599 & -5.26332 & 6.0803 & 0.0017 & 5226 \\
3209522618130302720 & 83.7368 & -5.19250 & 3.97728 & 0.00059 & 5020 \\
3209527634652049408 & 83.8477 & -5.18104 & 7.65863 & 0.00045 & 3928 \\
3209528012609165440 & 83.9128 & -5.14900 & 6.7452 & 0.0052 & 5199 \\
3209528523708699392 & 83.8055 & -5.15549 & 9.742 & 0.052 & 4412 \\
3209528905962364544 & 83.8595 & -5.14445 & 2.7453 & 0.0028 & 5946 \\
3216061001461614976 & 84.7167 & -2.77882 & 10.763 & 0.083 & 4510 \\
3216062272771926144 & 84.7216 & -2.73141 & 7.8142 & 0.0059 & 3931 \\
3216107490187638784 & 84.6136 & -2.75270 & 7.0805 & 0.0026 & 4332 \\
3216109861009561728 & 84.6843 & -2.67216 & 10.890 & 0.011 & 4623 \\
3216432464593272960 & 84.8872 & -2.79701 & 2.70295 & 0.00042 & 4460 \\
3216439023007610624 & 84.8857 & -2.66224 & 6.8943 & 0.0019 & 5105 \\
3216439847641330816 & 84.8550 & -2.63946 & 9.762 & 0.022 & 4405 \\
3216446547790307584 & 84.9493 & -2.54026 & 1.57340 & 0.00015 & 4934 \\
3216446689524900480 & 84.9687 & -2.53400 & 12.923 & 0.039 & 4736 \\
3216447239280699008 & 85.0141 & -2.48375 & 7.357 & 0.031 & 4505 \\
3216485722187722752 & 84.7985 & -2.60080 & 12.1119 & 0.0058 & 4488 \\
3216487749412405376 & 84.8453 & -2.55920 & 7.2427 & 0.0017 & 3760 \\
3216488329232163328 & 84.7725 & -2.54177 & 5.7223 & 0.0047 & 4678 \\
3216490223313304576 & 84.6478 & -2.53103 & 10.4049 & 0.0044 & 4824 \\
3216491151026233472 & 84.6678 & -2.50516 & 7.8721 & 0.0011 & 4208 \\
3216492869013787264 & 84.7002 & -2.45393 & 8.4758 & 0.0082 & 4875 \\
3216502798977562112 & 84.5281 & -2.50632 & 7.3507 & 0.0020 & 3875 \\
\hline
\end{longtable}

\bsp	
\label{lastpage}
\end{document}